\documentclass[printer]{aa}
\usepackage{graphicx}
\usepackage{txfonts}
\usepackage{bbm}

\begin{document}

\def\mxi{\mbox{\boldmath $\xi$}}
\def\mphi{\mbox{\boldmath $\phi$}}
\def\mkappa{\mbox{\boldmath $\kappa$}}
\def\bfe{{\bf e}}
\def\bfb{{\bf b}}
\def\bfu{{\bf u}}
\def\bfv{{\bf v}}
\def\N{{\cal N}}
\def\E{{\cal E}}
\def\XMM{{\it XMM-Newton}}

\title{Unbinned maximum-likelihood estimators for low-count data}
\subtitle{Applications to faint X-ray spectra in the Taurus Molecular Cloud}

\authorrunning{Arzner et al.}
\titlerunning{unbinned ML estimators}

\author{Kaspar Arzner\inst{1}, Manuel G\"udel\inst{1}, Kevin Briggs\inst{1}, Alessandra Telleschi\inst{1}, \\
	Manfred Schmidt\inst{2}, Marc Audard\inst{3}, Luigi Scelsi\inst{4}, \and Elena Franciosini\inst{4}}
\offprints{K. Arzner}
\institute{Paul Scherrer Institut, CH 5232 Villigen, Switzerland\\
              \email{arzner@astro.phys.ethz.ch}
	      \and
	      Zurich University of Applied Sciences, Winterthur, Switzerland
	      \and
	      Columbia Astrophysical Laboratory, 550 West 120th St, MC 5247, New York, NY 10027, USA
	      \and 
	      Dipartimento di Scienze Fisiche e Astronomiche, Piazza del Parlamento 1, 90134 Palermo, Italy
	      \and INAF-Osservatorio Astronomico di Palermo, Piazza del Parlamento 1, 90134 Palermo, Italy}

\abstract{
Traditional binned statistics such as $\chi^2$ suffer from information loss and arbitrariness of the binning 
procedure, which is especially important at low count rates as encountered in the \XMM~Extended
Survey of the Taurus Molecular Cloud (XEST). We point out that the 
underlying statistical quantity (the log likelihood $L$) does not require any binning beyond
the one implied by instrumental readout channels, and we propose to use it for low-count data.
The performance of $L$ in the model classification and point estimation problems is explored 
by Monte-Carlo simulations of {\it Chandra} and \XMM~X-ray spectra, and is compared to the 
performances of the binned Poisson statistic ($C$), Pearson's $\chi^2$ and Neyman's $\chi^2_N$, 
the Kolmogorov-Smirnov, and Kuiper's statistics. 
It is found that the unbinned log likelihood $L$ performs best with regard to the expected chi-square distance between 
true and estimated spectra, the chance of a successful identification among discrete candidate models,
the area under the receiver-operator curve of reduced (two-model) binary classification problems, 
and generally also with regard to the mean square errors of individual spectrum parameters.
The $\chi^2$ ($\chi^2_{\rm N}$) statistics should only be used if more than 10 (15) predicted counts 
per bin are available. From the practical point of view, the computational cost of evaluating $L$ 
is smaller than for any of the alternative methods if the forward model is specified in terms 
of a Poisson intensity and normalization is a free parameter. The maximum-$L$ method is applied to 14 XEST observations, 
and confidence regions are discussed. The unbinned results are compared to binned XSPEC results,
and found to generally agree, with exceptions explained by instability under re-binning and by
background fine structures. In particular, HO Tau is found by the unbinned method to be rather
cool ($kT$ $\sim$ 0.2 keV), which may be a sign of shock emission.
The maximum-$L$ method has no lower limit on the available counts, and allows to treat weak sources
which are beyond the means of binned methods.
}

\date{Received 30. January 2006; Accepted 6. September 2006}

\maketitle

\section{Introduction}

Energy histograms are often used as spectrum estimators. 
While this procedure has the advantage of simplicity, the grouping of counts into a 
histogram is also associated with information loss. This becomes especially important
if only few counts are available, so that the spectral fine structures are sparsely sampled
by the observed counts. Such a situation arises in the \XMM~Extended Survey of the Taurus Molecular Cloud 
(XEST; G\"udel et al. \cite{guedel06a}), thus prompting the search for alternative unbinned methods. 

Another motivation for histogram formation is to use $\chi^2$ as a simple and simply coded measure of agreement 
between theory and observation. The role of $\chi^2$ is thus to provide a plausibility ordering of
alternative models, and to assess their absolute credibility. Since the histograms follow a multinomial
distribution, which becomes gaussian only in the limit of infinite sample size, the use of
$\chi^2$ represents an approximation and corrections must be applied for small $n$
(e.g., Kendall \& Stuart \cite{kendall58}, Wachter \cite{wachter79}, Nousek \cite{nousek89}, Mighell \cite{mighell99}, 
Arzner \cite{arzner04}). Alternatively, one may use the binned Poisson ($C$) statistics.
There exists, however, a simpler solution.

Namely, the relevant statistical quantity, the likelihood function, can be defined for an unlimited 
instrumental resolution without any reference to binning. Using this unbinned expression avoids 
arbitrariness of histogram formation, and thus a potential source of discrepant results. It has been 
successfully applied to ROSAT (Boese \& Doebereiner \cite{boese01}) and EGRET (Digel \cite{digel00}) 
observations. Related approaches invoke piecewise-constant intensity models (Scargle 
\cite{scargle98}; see Stelzer et al. \cite{stelzer06} for an application to the XEST) 
or transformations to uniform null hypotheses (Kinoshita \cite{kinoshita02}), and
applications in other fields than astronomy include particle physics (Baker \& Cousin \cite{baker84}) 
and medicine (Miller et al. \cite{miller02}). While the $C$ statistic is often used in astronomical 
applications (e.g., Dolphin \cite{dolphin02} for star formation statistics; Babu and Feigelson \cite{babu96}
for a general overview), the unbinned (exact) Poisson likelihood is less often used. Alternative unbinned methods 
such as the Kolmogorov-Smirnov or Kuiper's statistics have the advantage of being (asymptotically) model-independent,
which, however, also entails sub-optimal performance if the modeling was correct.

In this article we revisit the issue by Monte-Carlo simulations of \XMM~and {\it Chandra} CCD
spectra of the Taurus Molecular Cloud (TMC). By simulating counts from a known spectrum, and applying different statistics 
in order to find the best-fit model, we assess the performances of the various statistics,
and thereby gain a more differentiated picture. The present study represents an extension of the work of
Nousek (\cite{nousek89}) and Wachter et al. (\cite{wachter79}) to more complex spectral models and a 
broader range of statistics.

The article is organized as follows. Section 2 introduces the models for sources and 
background spectra. Section 3 discusses the counting statistics.
Section 4 describes measures af agreement between models, and between models and data, which can be
used to assess the performance of various statistics. Section 5 describes the 
Monte-Carlo simulations and their numerical results. Real-data applications from the TMC are presented 
in Section 6, where also the issue of confidence regions is briefly touched,
and the unbinned estimates are compared to the binned estimates from the XSPEC software.
Section 7 contains a summary and conclusions.

\section{Spectral Models}

We start with introducing the spectral models $f(E)$ used in this study, which represent the
expected number of counts per unit energy.

In modern observations, $E$ can usually not assume continuous values but is
restricted to a discrete set of instrumental output channels. For example, the PN detector of \XMM~
has by default $N_c$ = 4096 such channels; {\it Chandra}/ACIS has $N_c$ = 1024. In both cases, 
the channel separation $\delta E$ is much smaller than the instrumental resolution as given by
the spectral response matrix, so that $f(E)$ is fully resolved. In order to stress the quasi-continuous
nature of the channel coverage, we shall write $f(E) \, dE$ rather than $f_j \, \delta E_j$ ($j=1...N_c$)
unless stated otherwise. The channels are also 
much finer than the bins typically used when computing the $C$ or $\chi^2$ statistics. In this sense,
methods which directly use the energy channels will be referred to as `unbinned', whereas
methods which group the channels first into larger `bins' will be referred to as `binned'.

In what follows we exclusively work with counts
and focus on the problem of finding the best spectral model for a given set of counts. Accordingly,
we do not consider here the problem of spectral deconvolution, but absorb the instrumental 
response in the forward model $f(E)$. The symbol $E$ thus denotes the observed (channel) energy rather 
than the incoming photon energy. 

\subsection{\label{source_sect}Source}

\begin{figure}[h]
\includegraphics[width=0.5\textwidth]{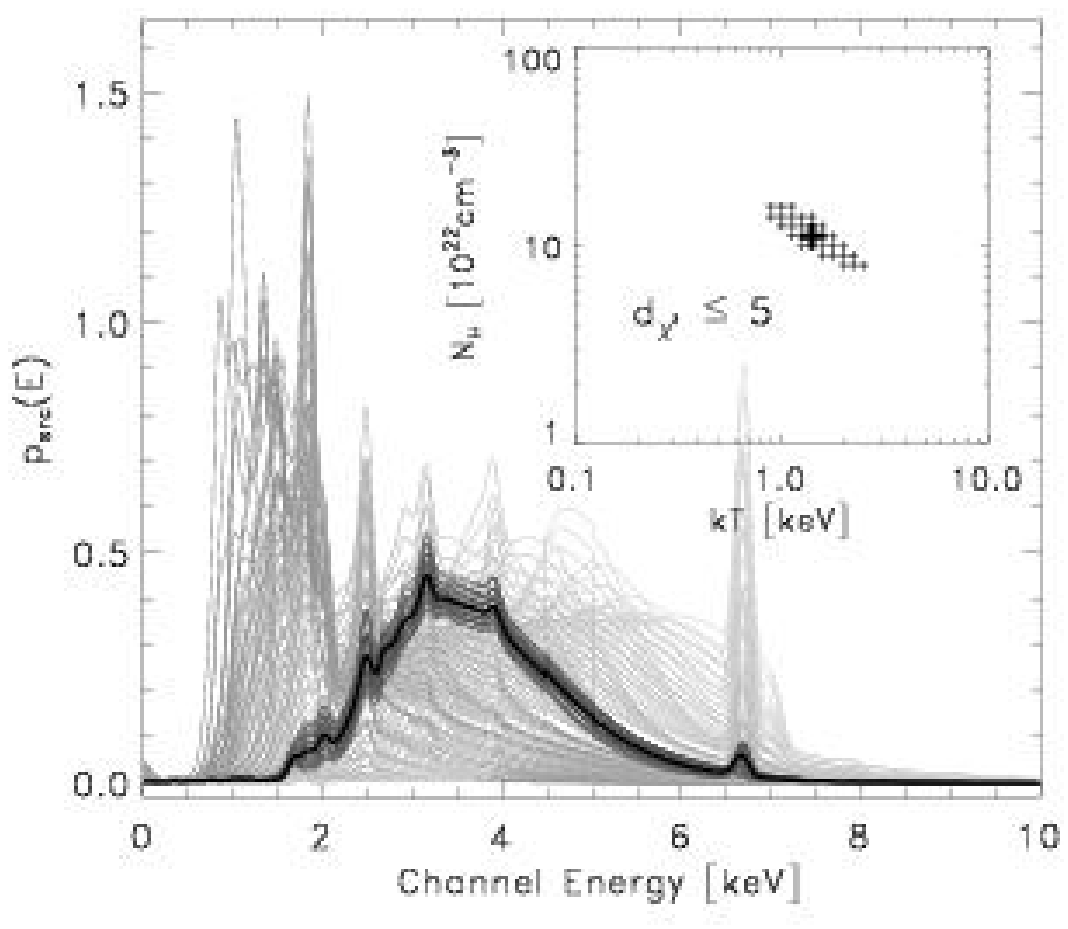}
\includegraphics[width=0.5\textwidth]{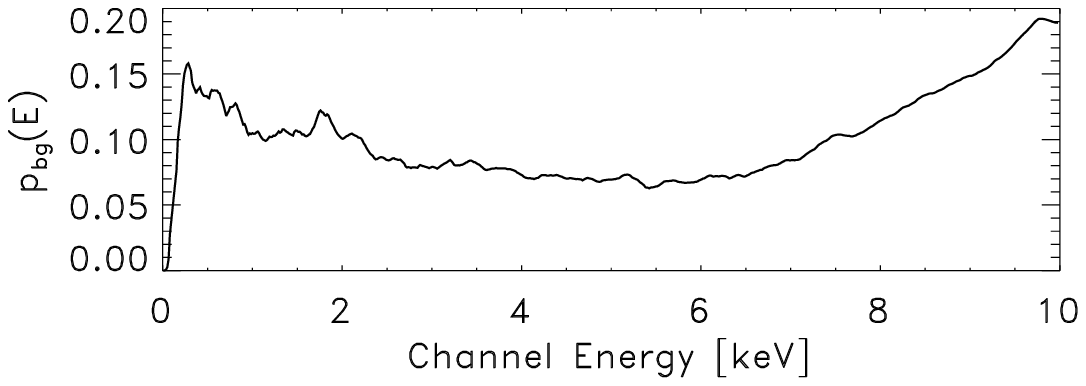}
\caption{Top: normalized source models used in this study. An example model (black) and its neighbours 
(dark gray) are highlighted for better clarity. Inlet: the example model and its neighbours
in parameter space $(kT,N_H)$. Bottom: normalized background model.}
\label{template_spectra_fig}
% created by ~manuel/unbinned/prog/show_template_spectra.pro and convert_chandra_background.pro
\end{figure}

Our source models assume absorbed collisional ionization equilibrium 
(APEC; Smith et al. \cite{smith01}, as implemented in XSPEC; Arnaud \cite{arnaud96}), convolved
with the instrumental response. The templates are defined on the $N_c$ instrumental channels, and
depend on only two physical parameters: the (electron) temperature $kT$ in units of keV, and the hydrogen
column density $N_H$ in units of 10$^{22}$ cm$^{-2}$. This simple model is adapted to the faint sources
addressed by the $L$ statistic. The parameters $(kT,N_H)$ are specified on a double logarithmic lattice with 
sufficiently dense coverage that intermediate spectra may be interpolated with error $d_{\chi^2} < 10^{-5}$ 
(see Sect. \ref{dchi2_sect}). This gridding is used for simulations only.
%
% see arzner@psw320:/afs/psi.ch/user/a/arzner/manuel/unbinned/prog/check_interpol_err.pro
%
The total number of expected counts $\N_{\rm src}$ is an additional parameter. 
The normalized templates are displayed in Fig. \ref{template_spectra_fig} (top panel, selection).
For better physical clarity, the $x$-axis refers to channel energy rather than to channel number.
In the Monte-Carlo simulations, atomic lines are included assuming a fixed 
metallicity of 0.2 times the solar abundances of Anders \& Grevesse (\cite{anders89}), and we use the {\it Chandra} 
rather than the \XMM~instrumental response because of its smaller number of instrumental channels, 
which accelerates the simulations. This choice applies to the simulations only, and we return to the \XMM~
instrumental response and to a more refined abundance model when dealing with real XEST data. The function 
$f(E)$ contains a background which is interpolated from calibration
observations (see Sect. \ref{bkg_section}). The astrophysical relevance of the parameters ($kT,N_H,\N$) is elucidated in the 
accompanying article of G\"udel et al. (\cite{guedel06a}). From an empirical point of view, the 
effect of large $N_H$  is to suppress (absorb) the spectrum from at low energies, and the effect of large $kT$ 
is to enhance it mostly at high energies.

\subsection{\label{bkg_section}Background}

Although {\it Chandra}, owing to its high spatial resolution, has little background, the present simulations 
include an additive background for the sake of generality and applicability to \XMM~data. The background is included 
in the forward model according to 

\begin{eqnarray}
f(E) 	& = & f_{\rm src}(E) + f_{\rm bg}(E) \label{f_tot} \\
\N& = & \N_{\rm src} + \N_{\rm bg} \, . \label{N_tot}
\end{eqnarray}

The background model $f_{\rm bg}(E)$ is obtained from an extended source-free extraction region. Since the
background estimate is typically much better than the source estimate, the background is 
considered as known; this represents a simplifying assumption.
The background density\footnote{For numerical conciseness, we use here channel-indexed ($f_j \, \delta E_j$)
rather than continuous ($f(E) \, dE$) to  notation.} 
$f_{{\rm bg},j}$ is determined from the observed background histogram
$H_{{\rm bg},j}$ by minimizing the goal function $\sum_{j=1}^{N_c} [f_{{\rm bg},j}-H_{{\rm bg},j}]^2 + \sum_{j=1}^{N_c-1} 
(\sigma(E_j)/\delta E_j)^2 [f_{{\rm bg},j+1}-f_{{\rm bg},j}]^2$,
where $\sigma(E)$ = $\kappa \sqrt{E/{\rm keV}}$ represents the approximate instrumental resolution.
The coefficient $\kappa$ is 0.07 for {\it Chandra}/ACIS, and 0.05 for \XMM/EPIC, and is chosen such
that $\sigma(E)$ gives the full width half maximum energy resolution of the respective instruments.
The choice of the above goal function amounts to first-order regularization and yields a tridiagonal system
which is easily solved numerically. It ensures\footnote{
These properties are most easily seen from a continuous version 
where $\int_B [(f_{\rm bg}(E)-H(E))^2+\sigma^2(E) f'_{\rm bg}(E)^2] \, dE$ is 
minimized with respect to $f_{\rm bg}(E)$. Carrying out the variation (with $f_{\rm bg}(E)$ kept fixed at the boundaries) one 
obtains $f_{\rm bg}(E)-(\sigma^2(E) f'_{\rm bg}(E))'=H(E)$. This may be integrated to $\int f_{\rm bg}(E) dE = \int H(E) dE$,
assuming that $f'(E)$ vanishes at the boundaries. Multiplying the above differential equation by $(E-E_k)^2$, taking
$H(E) = \delta(E-E_k)$, and assuming that $\sigma^2(E)$ varies slower than $f_{\rm bg}(E)$, 
one finds $\int f_{\rm bg}(E) (E-E_k)^2 dE \simeq 2 \sigma^2(E_k) \int f_{\rm bg}(E) dE$ by integration by parts, q.e.d.}
that $f_{{\rm bg},j}$ is unbiased in the (weak) sense that
$\sum_j f_{{\rm bg},j} = \sum_j H_{{\rm bg},j}$, and that $f_{{\rm bg},j}$ has the approximately correct energy resolution 
in the sense that the response $f_{{\rm bg},j}^*$ to a unit pulse $H_{{\rm bg},j} = \delta_{j,k}$ has the mean-square 
width $\sum_j (E_j-E_k)^2 f_{{\rm bg},j}^*\big/ \sum_j f_{{\rm bg},j}^* \simeq 2 \sigma^2(E_k)$.

To summarize, we use a smooth density estimator for the background which interpolates the observed
background counts, preserves their total number, and has the (approximately) correct instrumental resolution.
The normalized background spectrum of a typical {\it Chandra}/ACIS observation 
is shown in Fig. \ref{template_spectra_fig} (bottom panel). The peak at 1.8 keV is due to silicon in the CCD.

\section{\label{counting_sect}Counting statistics}

Photon counting experiments such as \XMM~and Chandra follow the Poisson statistics, which is briefly recalled here.
More detailed discussions may be found e.g. in Feller (\cite{feller68}), Eadie et al. (\cite{eadie71}), Santal\'o (\cite{santalo76}), 
Wachter (\cite{wachter79}), Reiss (\cite{reiss93}), and Protassov et al. (\cite{protassov02}).

\subsection{The Poisson process}

We consider throughout a non-homogeneous Poisson process in an interval $B = (E_{\rm min}, E_{\rm max})$ of the real axis
with finite intensity $f(E)$. Generalizations to (pointwise) infinite $f(E)$ and higher dimensions can be found e.g. 
in Reiss (\cite{reiss93}). The Poisson process is then characterized by the following two properties: (i) 
the probability of finding $n_j$ counts in an sub-interval $\Delta_j \subset B$ is

\begin{equation}
{\rm Prob}(n_j) = \frac{e^{-\lambda_j} \lambda_j^{n_j}}{n_j!} \;\;\; \mbox{where} \;\;\; \lambda_j = \int_{\Delta_j} f(E) dE \, ,
\label{poisson}
\end{equation}

and (ii) the numbers of counts in disjoint $\Delta_j$'s are statistically independent of each other.
The first Equation in (\ref{poisson}) defines the Poisson distribution, and $\lambda_j$ is called the
parameter of the Poisson distribution.

When counts from different intensity functions $f_k(E)$ are added, the
resulting process is again Poissonian with intensity $\sum_k f_k(E)$. This fact may
be used to include an observational background, and to construct a Poisson process of
arbitrary overall expectation

\begin{equation}
\N = \int_B f(E) \, dE
\label{nrm}
\end{equation}

by first drawing the number $n$ of counts from a Poisson distribution (Eq. \ref{poisson})
with parameter $\N$, and then drawing each count $E_i$, $i = 1 \, ... \,n$ from the probability density 

\begin{equation}
p(E) = f(E)/\N
\label{p(E)}
\end{equation}

(see Reiss \cite{reiss93}, Theorem 1.2.1). The above construction relies on the
factorization of the number of counts from their position on the energy axis.
The integral in Eq. (\ref{nrm}) is numerically approximated as a sum over instrumental channels.

\subsection{\label{L_sect}Likelihood function and asymptotic forms}

The likelihood function is defined as the probability of finding a certain observation given the true model
from which the observation derives. Given the observed sample size $n$, the likelihood of 
the observation $\{ E_1, ..., E_n \}$ is
\begin{equation}
P(E_1, ... , E_n \, | \, kT, N_H) = p(E_1) \cdot ... \cdot p(E_n) \, . \label{L1}
\end{equation}
Equation (\ref{L1}) represents a probability density at $(E_1, \, ... \,, E_n)$. 
The shape of the function $p(E)$ is determined by the parameters $kT$ and $N_H$.

As outlined above, the normalization $\N$ factorizes out, and can be estimated from the total 
number of observed counts alone. We shall do this using the maximum-likelihood estimator (i.e., that
$\N$ which maximizes $e^{-\N}\N^n/n!$). Since the background contribution $\N_{\rm bg}$ is
known (Sect. \ref{bkg_section}), only the source contribution $\N_{\rm src}$ needs to be estimated;
this will be done using the maximum-likelihood estimator
\begin{equation}
\N_{\rm src} = \max (0, n-\N_{\rm bg}) \, .
\label{nrm_ML}
\end{equation}

The exact likelihood (Eq. \ref{L1}) can be recast in alternative and asymptotic forms which are
often used in astrophysics. When binned into finite bins $\Delta_j$ of predicted content $p_j = \int_{\Delta_j} p(E) \, dE$
containing $n_j$ observed counts, the likelihood (given the total number of observed counts $ \sum_{j=1}^{N_b} n_j = n$)
becomes the binomial distribution
\begin{equation}
P(n_1, ..., n_{N_b} | kT, N_H) = \frac{n!}{n_1! \cdot ... \cdot n_{N_b}!} p_1^{n_1} \cdot ... \cdot p_{N_b}^{n_{N_b}} \, .
\label{L2}
\end{equation}
In the asymptotic limit ($n_j \to \infty$), the standardized variables $(n_j-np_j)/\sqrt{np_j}$ become
normal and the quantity $X^2 = \sum_j x_j^2$ becomes $\chi^2$ distributed with $(n-1)$ degrees of freedom
(Kendall \cite{kendall58}). Accordingly, the log likelihood (logarithm of Eq. \ref{L2}) becomes chi square
distributed with ($n-1$) degrees of freedom, and we may use the chi square statistic. The approximation
of the logarithm of a multinomial by a chi square distribution is especially good when all the $n p_j$ are (approximately)
equal; in this case (and only in this case), the $n p_j$ need not be large by themselves (Wise \cite{wise63}).
There are different forms of the chi square statistics, and we shall return to these -- and to
alternative statistics -- in Sect. \ref{dist_stat_sect}.

\subsection{\label{bin_method_sect}Binning}

The choice of optimal bins (Schott \cite{scott92}) is an important issue which affects the 
outcome of the binned methods. Histograms can have uniform or non-uniform bins. While uniform
(equal-size) bins have the advantage of being independent of the data, they
are also less adapted and may contain very few counts. Non-uniform bins are usually chosen
to contain the same (predicted) probability mass or observed counts.

For either (uniform or non-uniform) type of bins, one needs to decide on the number of bins or
the average number of counts per bin. A well-known
method to choose the number $N_b$ of uniform bins is Sturges' rule (Sturges \cite{sturges26})

\begin{equation}
N_b = 1 + \log_2 n
\label{sturges}
\end{equation}

where $n$ is the total number of observed counts. Sturges' rule is based on the assumption that the
counts $E_i$ follow a normal distribution, and tends to under-estimate the number of bins needed to resolve more
complicated forms such as the $f(E)$ considered here. An alternative approach, which is more adapted to the 
actual shape of $p(E)$, minimizes the asymptotic mean square integrated error (AMISE). This implies a trade-off between
the integrated variance (due to finite counts per bin) and the integrated squared bias (due to variation of $p(E)$ 
across the bins). Expanding $p(E)$ to first order in each histogram bin, this
yields the expression

\begin{equation}
h_* = (6/R(p'))^{1/3} n^{-1/3}
\label{AMISE}
\end{equation}

for the optimal bin width, where $R(p') = \int_B p'(E)^2 dE$ measures the roughness of the spectrum 
(Scott \cite{scott92}, theorem 3.3).

In our simulations, the optimal number $n^*$ of counts per bin is either taken according to Eq. (\ref{sturges})
as $n^* = n/(1+\log_2 n)$, or according to Eq. (\ref{AMISE}) as $n^* = n^{2/3} B^{-1} (6/\langle R(p') \rangle)^{1/3}$
where the (unknown) true $R(p')$ has been replaced by an average $\langle R(p') \rangle$ over all models under consideration.
The choices based on Eqs. (\ref{sturges}) and (\ref{AMISE}) will be referred to as Sturges' and AMISE methods.
Alternatively, $n^*$ is fixed at a given value, similar as in the standard \XMM~and Chandra data analysis software.
This will be referred to as fixed-$n^*$ method; a typical choice is 10 counts per bin. (In XSPEC,
this becomes $\ge 10$ cts/bin for high-intensity flares which are not considered here.) For the problem
considered here ($\langle R(p') \rangle = 2.9$ cts/keV$^3$; $\N$ ranging from a few 10 to a few 100), the AMISE method 
gives less than 10 counts per bins, whereas Sturges' method gives a few to a few ten counts per bin.

Once the number of counts per bin $n^*$ is chosen,
the binning itself is performed using either uniform or non-uniform bins. For the case of non-uniform bins we use by default bins with equal
numbers of observed counts, with the $k$-th bin running from $E_{k n^*}$ to $E_{(k+1)n^*}$. This simple method is not optimal
but adopted for consistency with the standard analysis software. Alternative methods have also been considered, which are 
based on equal predicted bin content, and found to yield similar results.

\section{\label{fwd_model_sect}Distance measures}

Next we define measures of agreement between different models, and between models and observations.
These will allow to estimate models from the data, and to assess the discrepancy between
true and estimated models in the subsequent Monte-Carlo simulations.

\subsection{\label{dchi2_sect}Chi-square distance between models}

Different sets of parameters $(kT,N_H,\N)$ result in different model spectra, whose
discrepancy can be measured by a suitable distance in the space of predicted spectra ($f$-space). 
It is important to realize that it is the distance in $f$-space which is relevant for the observational
discrimination between alternative models, and not the (say, Euclidean) distance in 
parameter space, because the latter may degenerate solely due to ambiguous parameterization.
Recalling that $f(E)$ is the intensity of a Poisson process, and assuming that $f_1(E)$ is the true and
$f_2(E)$ is the estimated spectrum, we use here the chi square distance (e.g., Gibbs \cite{gibbs02})

\begin{equation}
d_{\chi^2}(f_1,f_2) = \int_B \frac{(f_1-f_2)^2}{f_1} \, dE \, .
\label{d_chi2}
\end{equation}

An illustration of $d_{\chi^2}$ is given in Fig. \ref{template_spectra_fig}, where the dark gray spectra 
deviate by $d_{\chi^2} \le 5$ from the reference spectrum (black), assuming that $\N$ = 100. The
parameters $kT$ and $N_H$ are not independent, as indicated by the banana-shaped region
in Fig. \ref{template_spectra_fig} (inlet). It should be pointed out that it is {\it not} this kind of 
degeneracy which is addressed in the present paper, but rather the principal capability of various
statistics to distinguish between models which differ in the sense of Eq. (\ref{d_chi2}). 
Note that $d_{\chi^2}(f_1,f_2)$ is not a metric since it is not symmetric in its arguments,
and also violates the triangle inequality with respect to the first argument $f_1$ (but not 
with respect to the second argument $f_2$).
%
% see ~/manuel/unbinned/prog/check_triangle.pro
%
Alternative probabilistic distances may be found e.g. in Basseville (\cite{basseville89}), Rachev (\cite{rachev91}),
Reiss \cite{reiss93}, M\"uller (\cite{mueller95}), O'Sullivan et al. (\cite{osullivan98}), Robinson et al. (\cite{robinson00}),
Johnson \& Sinanovic (\cite{johnson01}), and Gibbs \& Su (\cite{gibbs02}). We have chosen here Eq. (\ref{d_chi2}) 
because it does not require probabilistic normalization of $f$, and because it is not explicitly adapted to the $L$ 
statistics, the performance of which is to be demonstrated.

\subsection{\label{dist_stat_sect}Distance between models and measurements}

After having defined a measure of agreement between different models, we shall specify measures
of agreement between models and observations. To this end, we consider the unbinned likelihood
(Eq. \ref{L1}) together with a selection of some of the statistics
which are often used in an astrophysical and astronomical context (e.g., Gosset \cite{gosset87}; 
Wachter et al. \cite{wachter79}; Nousek \& Shue \cite{nousek89};
Babu \& Feigelson \cite{babu96}; Mighell \cite{mighell99}; Metchev \cite{metchev02};  Paltani \cite{paltani04}):

\begin{eqnarray}
L 	& = & \sum_{i=1}^n \ln p(E_i) \label{L} \\
C	& = & \sum_{j=1}^{N_b} (n_j \ln \lambda_j - \lambda_j) \;\;\; \mbox{with} \;\;\; 
		\lambda_j = \int_{\Delta_j} \hspace{-1mm} dE \, f(E) \label{C} \\
\chi^2  & = & \sum_{j=1}^{N_b} \frac{(n_j - \lambda_{j})^2}{\lambda_{j}} \label{chi2} \\
\chi^2_{\rm N}  & = & \sum_{j=1}^{N_b} \frac{(n_j - \lambda_{j})^2}{n_{j}} \;\;\;\;\;\;\; (n_j \not= 0) \label{chi2_Neyman} \\
D	& = & \max_E | P(E) - S_N(E)| \;\; \mbox{with} \;\; P(E) = \int_{E_{\rm min}}^E \hspace{-1mm} p(E') \, dE' \label{KS} \\
V	& = & \max_E \big(P(E)-S_N(E)\big) + \max_E \big(S_N(E)-P(E)\big) \, . \label{V}
\end{eqnarray}

Above, $L$ is the unbinned log likelihood (logarithm of Eq. \ref{L1}), $C$ is the $C$ statistic (Cash \cite{cash79}),
and $D$ is the Kolmogorov-Smirnov (or uniform) distance between two probability densities, of which the Kuiper 
statistic $V$ is a variant with more balanced sensitivity across $[E_{\min},E_{\max}]$ (Kuiper \cite{kuiper62}).
$P(E)$ is the predicted cumulative distribution, and $S_N(E)$ is its observed counterpart, i.e., 
$S_N(E)$ is the number of counts with energy smaller than $E$, normalized by the total number $n$ of counts 
(thus $0 \le S_N(E) \le 1$). $N_b$ denotes the number of bins of the binned methods (Sect. \ref{bin_method_sect}).
The subscript `N' in Eq. (\ref{chi2_Neyman}) stands for Neyman; it is undefined for bins with $n_j = 0$ which
are therefore excluded. (The reference to Neyman is for his detailed analysis; the replacement of
theoretical variances by observed ones has already been proposed by Pearson -- see Hald \cite{hald98}).
Note that evaluation of $L$ is computationally less expensive than evaluation of the binned statistics when
there are fewer counts than channels and if the model provides $p(E)$ rather than $P(E)$.

It should be pointed out that the selection of statistics (Eqs. \ref{L}-\ref{V}) is aimed at our astrophysical 
application of estimating X-ray spectra, and is not free of personal bias.
The $L$ statistic is the one which is primarily addressed by the present article.
The $C$ and $\chi^2_{\rm N}$ statistics are implemented in the standard XSPEC software pakage; they are included 
for benchmarking and comparison of observational results. The $\chi^2$ statistic is an obvious variant of $\chi^2_{\rm N}$, 
which is obtained as the asymptotic limit of the binned Poisson log likelihood (cf. Sect. \ref{L_sect}).
Among the unbinned statistics, the Kolmogorov-Smirnov $D$ is probably the most fundamental one 
used in astrophysics. The Kuipers statistic $V$ has been included as an example of an improved version of $D$.
Others (such as the Anderson-Darling statistic) could have been included as well, but we decided to restrict 
the list in order not to overload the diagnostics.

In order to treat all statistics (\ref{L}) - (\ref{V}) on the same footing, the source normalization $\N_{\rm src}$ 
is always estimated from Eq. (\ref{nrm_ML}), and only the shape parameters ($kT,N_H$) are estimated differently for
each statistic. The case $\N_{\rm src} = 0$ is rarely met in practice.

\begin{figure}[h]
\includegraphics[width=0.45\textwidth]{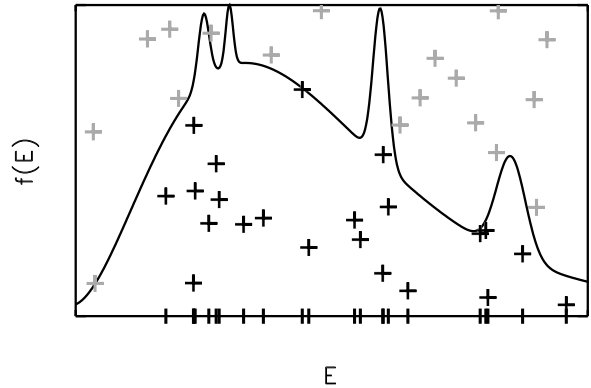}
\caption{Schematic construction of a continuous non-homogeneous Poisson process by the rejection method:
out of $N_p$ uniform random points in the rectangle $B \times [0,\max (f)]$ 
only those beneath the curve $f(E)$ are accepted; their abscissae constitute the event list (ticks). 
The number $N_p$ itself is Poisson distributed with expectation value $|B| \times \max (f)$.}
\label{poisson_sample_fig}
% created by ~manuel/ISSI/show_poisson_sample.pro
\end{figure}

\begin{figure}[h]
\includegraphics[width=0.45\textwidth,height=0.42\textwidth]{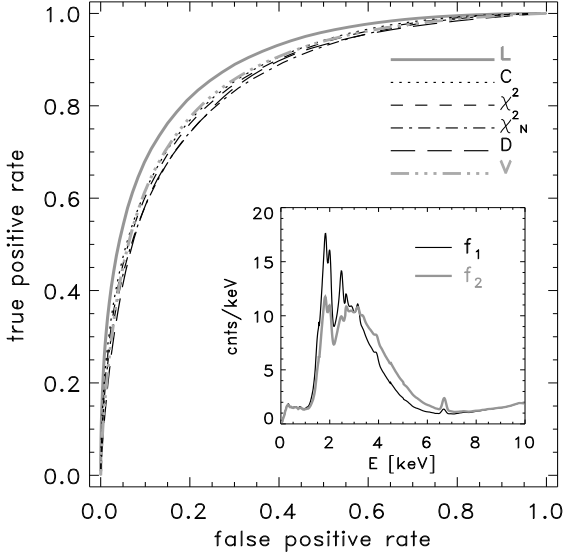}
\caption{\label{roc_fig}Receiver-operator characteristics for the binary classification
problem with only two models (inlet).}
% created by ~unbinned/prog/revised/main5.f90 and result5.pro
\end{figure}

\subsection{\label{ROC_sect}Receiver-Operator Characteristics}

Equation (\ref{d_chi2}) can be used to measure the discrepancy between true and estimated models, and thus to
assess the performance of the statistics (Eqs. \ref{L} - \ref{V}). Alternatively, their performance 
can also be characterized in terms of the classical receiver-operator characteristics (ROC; Peterson et al. 
\cite{peterson54}; Hanley \& McNeill \cite{hanley82}; Michel \cite{michel03}). 
This diagnostics applies to the binary classification problem, and 
visualizes the trade-off between the reliability and completeness of detections.
In its standard formulation, the binary classification problem considers two types of objects (healthy and ill patients, 
for example) with a real-valued attribute. The attribute is a random variate, the
probability density of which depends on the object type. The task is to classify the objects by their attributes, which is
most simply done by applying a threshold. The classification is only unique if the probability densities of the attribute
do not overlap. There are four possible outcomes of the classification procedure,
with frequencies indicated in brackets: a type 1 object may be correctly classified as type 1 ($n_{11}$) 
or erroneously as type 2 ($n_{12}$), and a type 2 object may be correctly classified as type 2 ($n_{22}$) 
or erroneously as type 1 ($n_{21}$). From these frequencies we define the sensitivity $n_{11}/(n_{11}+n_{12})$
and the specificity $n_{22}/(n_{21}+n_{22})$. (The asymmetric naming stems from associating type 1 and 2
with asymmetric entities like healthy and ill patients.) Let us assume that a low attribute
is typical for type 1, and a high attribute is typical for type 2. If the threshold is high, only few
objects are accepted, but most of them are correctly classified as type 2. However, 
the bulk of type 2 objects is missed since it interferes with type 1.
Thus the specificity is high but the sensitivity is small. Conversely, a low threshold detects all type 2
objects, but at the expense of type 1 contamination. In this case, the sensitivity is high but
the specificity is low. As the threshold is varied, it traces out a trade-off between specificity and sensitivity that
is traditionally displayed in a (1-specificity) versus sensitivity diagram, equivalent
to false positive rate versus true positive rate. This graph is called the receiver-operator characteristics (ROC).

In order to apply the ROC procedure to the \XMM~spectrum estimation problem, we restrict the latter to only
two possible models $f_1$ and $f_2$, and use the log likelihood ratio $L_2 - L_1$ as an attribute, and
similarly $C_2-C_1$, $\chi^2_1-\chi^2_2$, $\chi^2_{\rm N,1} - \chi^2_{\rm N,2}$, $D_1-D_2$, and $V_1-V_2$.
By convention, a low attribute thus hints to model 1 and a high attribute to model 2. We then choose at random
one of the two models, generate an eventlist, and compute the above attributes. This
procedure is repeated some $10^5$ (=$\sum_{ij}n_{ij}$) times, and the sensitivity and specificity are 
computed for varying thresholds. The resulting ROC curves are shown in Fig. \ref{roc_fig}. The two models
$f_1$ and $f_2$ have the same hydrogen column density $N_H$ = 4.5 and normalization $\N$ = 30 but their temperatures
differ by a factor two, $kT_1$ = 1.4 keV and $kT_2$ = 2.5 keV. Such a difference is within astrophysical expectations.
The chi square distance between the two models is $d_{\chi^2}(f_1,f_2) = 3.2$. Different curves represent
different statistics. Ideally (100\% sensitivity and 
100\% specificity), the ROC curve would go through the upper left corner (0,1) of Fig. \ref{roc_fig}, whereas a completely
non-discriminating test would yield a straight line along the diagonal from (0,0) to (1,1). The actually
realized curves are between the two extremes, with the exact likelihood coming closest
to the ideal point (0,1). 

The ROC curve can be used to assess the performance of the statistics (Eqs. \ref{L} - \ref{V}) in the  
binary classification problem. A commonly used performance measure is the area under the ROC curve (AUC).
Like the chi square distance, the AUC gives a measure of observable discrepancy between two models.
For Fig. \ref{roc_fig} we obtain the AUC's 0.761 ($L$), 0.705 ($C$), 0.679 ($\chi^2$),
0.655 (${\chi^2_N}$), 0.699 ($D$), and 0.727 ($V$), thus establishing the ordering 
$L \succ V \succ C \succsim D \succ \chi^2 \succ \chi^2_{\rm N}$.

\section{Monte-Carlo simulations}

In order to explore the performance of the unbinned likelihood and compare it to the alternative statistics
we have conducted extended Monte-Carlo simulations. 
The simulations involve two general steps. In the first step, a model is chosen at random and an event list is generated
by either the inversion or rejection method (Devroye \cite{devroye86}). The inversion method
is described in Appendix A, and the rejection method is illustrated in Fig. \ref{poisson_sample_fig}. 
The inversion method is faster and thus used in most simulations; both methods have been checked against
the analytical Poisson probabilities.
% see ~/manuel/unbinned/prog/check_invmethod.pro
% see ~/manuel/unbinned/prog/check_rejmethod.pro
In the second step, we compare the event list with models which are close enough to the true model
in a $d_{\chi^2}$ sense in order to be potentially successful candidates. This is done using all statistics listed 
in Eqns. (\ref{L}) - (\ref{V}); the decision rule is to accept the model with largest ($L, C$) or 
smallest ($\chi^2, \chi^2_{\rm N}, D, V$). The above two-step procedure is repeated for many realizations of the 
event list, and several diagnostics are applied. When the parameter $\N_{\rm src}$
is varied in the simulations, this is done by taking it uniformly 
distributed out of the indicated interval. The background $\N_{\rm bg}$ is kept fixed. Several runs 
addressing different aspects have been performed.

\begin{figure}[h]
\centerline{
\includegraphics[width=0.24\textwidth]{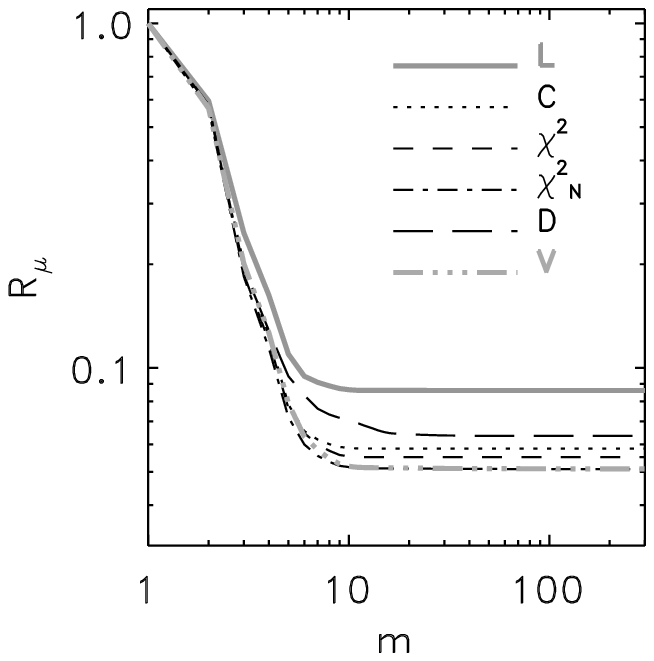}
\includegraphics[width=0.24\textwidth]{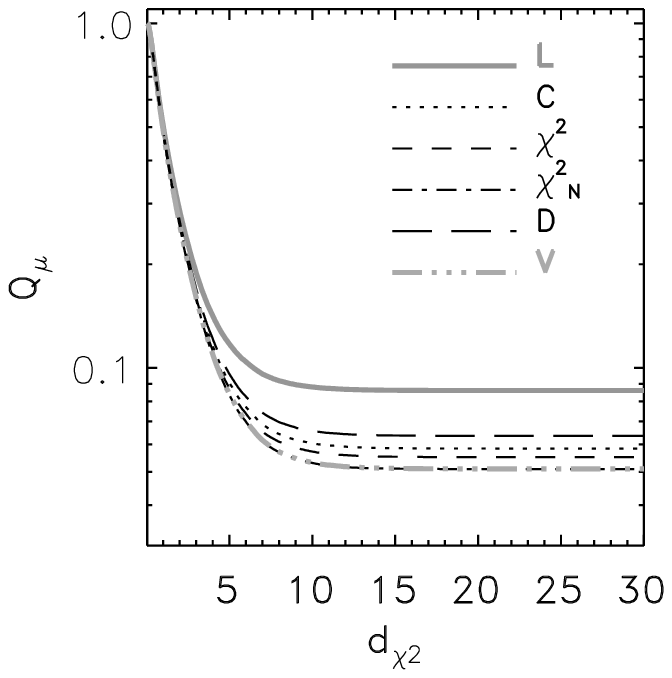}
}
\caption{\label{S_mu_sim_fig}The probability of successful identifications versus the number of 
$d_{\chi^2}$-ordered candidate models (left), and versus the maximal distance of the candidates
from the true model (right). See text. The simulation has $10 < \N_{\rm src} < 100$, $\N_{\rm bg}$ = 0, 
and the binned methods use uqual-size bins based on Sturges' rule, resulting on average in
$\langle n^* \rangle$ = 9.1 counts per bin.}
% created by ~unbinned/prog/revised/main1.f90 and result1.pro
\end{figure}

\subsection{\label{safe_discrim_sect}Safe discrimination}

In a first family of simulations, we investigate the minimum chi-square distance between two models which allows a safe distinction
on grounds of the observed event list. 

To this end, we work with the discrete set of models of Sect. \ref{source_sect} and call an estimate a
`successful identification' if the estimated model equals the true model. We are interested in the chance of a
successful identification, and proceed as follows. For each realization $\E$ of the eventlist drawn from a true model $f_1$,
the candidate models $\{ f_2 \}$ are ordered by increasing $d_{\chi^2}(f_1,f_2)$, and examined sequentially on grounds of the 
statistics (\ref{L}) - (\ref{V}). The first candidate is always taken as the true model. 
As the number $m$ of candidates increases, the initial 
(successful) estimate may be abandoned in favour of a wrong estimate. Let us assume that this happens, 
under statistics $\mu$ and for the realization $\E$, at the $k$-th candidate and define $\theta_\mu^\E(m)$ to be one 
for $m$ = 1 ... $k$ and zero for $m > k$. When the simulation is repeated for a large number of
realizations $\E$ (using different $f_1$), the quantity $R_\mu(m) = S_\mu(m)/S_\mu(1)$ with
$S_\mu(m) = \sum_\E \theta_\mu^\E(m)$ gives the probability of a successful identification in a search 
over $m$ $d_{\chi^2}$-ordered candidates using statistics $\mu$. 

An example of $R_\mu(m)$ is shown in Fig. \ref{S_mu_sim_fig} (left), with different curves referring to different 
statistics $\mu$. The binned methods invoked Sturges' rule to determine the number of counts per (equal-size) bin.
As increasingly unlikely candidates become included, the $R_\mu(m)$ converge to a constant 
values. This holds true for all statistics; however, there is a clear performance
ordering $R_L \succ R_D \succ R_C \succ R_{\chi^2} \succ R_V \succsim R_{\chi^2_{\rm N}}$.
The exact likelihood performs best, followed by the Kolmogorov-Smirnov distance.

While the quantity $R_\mu(m)$ directly relates to the simulation procedure, the number $m$ by itself is not
of interest and should be replaced by $d_{\chi^2}$ in order to obtain a more meaningful characteristic.
This is achieved by replacing the order indicator $\theta_\mu^\E(m)$ by a distance indicator $\theta_\mu^\E(d_{\chi^2})$
which switches from unity to zero at the distance of the first erroneously accepted candidate,
and proceeding in the same way as for $R_\mu(m)$. As a result we obtain the probability $Q_\mu(d_{\chi^2})$ 
of a successful identification in a $d_{\chi^2}$-ordered search among (discrete) candidates with maximal distance $d_{\chi^2}$ 
from the true model. The graphs of $Q_\mu(d_{\chi^2})$ are shown in the right pannel of Fig. \ref{S_mu_sim_fig},
referring to the same simulation as Fig. \ref{S_mu_sim_fig} (left).
Like $R_\mu(m)$, $Q_\mu(d_{\rm \chi^2})$ converges at large $d_{\chi^2}$ to a constant value which depends only on the 
statistics used. The performance ordering derived from $Q_\mu(d_{\chi^2})$ is the same as for $R_\mu(m)$. 

\begin{figure}[h]
\includegraphics[width=0.5\textwidth]{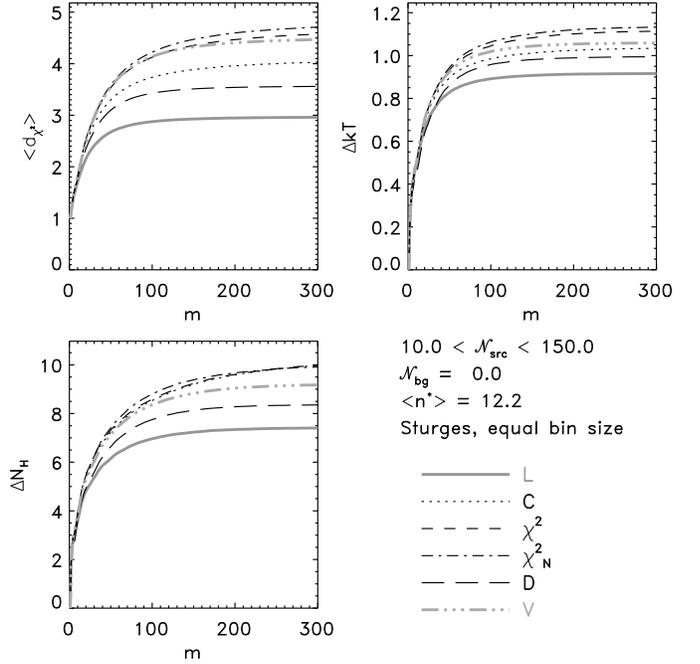}
\caption{\label{explore_d_diam_fig}Average $f$-space (top left) and parameter (other panels) deviations
between true and estimated models.
% for 50 $< \N <$ 150 and $\N_{\rm bg}$ = 0. 
The number $m$ of candidate models delineates the tested {\it volume} in
parameter space.}
% created by ~unbinned/prog/unbinned1.f90 and result1.pro
\end{figure}

Aside from the success rates one may ask for the error in parameter- and $f$-space if the identification
fails. The closer the estimated solution is to the true one, the better the method. 
Figure \ref{explore_d_diam_fig} displays the residuals in $f$- and parameter space for the same
situation as in Fig. \ref{S_mu_sim_fig} (left); the quantities ($\Delta kT, \Delta N_H$) are 
the standard deviations between true and estimated model parameters. They give the typical accuracy of best-fit 
parameters, averaged over the whole parameter space. As can be seen from Fig. \ref{explore_d_diam_fig}, 
the exact likelihood yields the smallest $\langle d_{\chi^2} \rangle$ (top left) and also
the smallest parameter errors ($\Delta kT, \Delta N_H$). The error in the normalization estimate
(Eq. \ref{nrm_ML}) is simply $\Delta \N = \sqrt{N}$ and is not shown. 

\begin{figure}[h]
\includegraphics[width=0.5\textwidth]{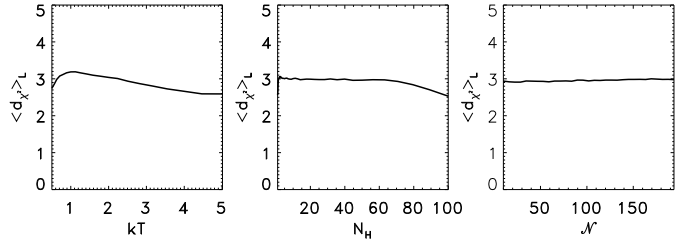}
\caption{\label{dchi2_vs_p1_fig}Average chi-square distance between true and estimated models as a 
function of the true model parameters (marginal distributions).}
% created by ~unbinned/prog/revised/main3.f90 and result3a.pro
\end{figure}

The stabilization of the curves observed in Figs. \ref{S_mu_sim_fig} (right) and \ref{explore_d_diam_fig} (top left) 
can be used to define a maximum chi-square distance $d_{\chi^2}^{\rm max}$ between true and estimated models above which 
mis-estimation becomes highly unlikely. From Fig. \ref{explore_d_diam_fig} (top left) 
we find that $\langle d_{\chi^2} \rangle_L$ stabilizes around 3, whereas Fig. \ref{S_mu_sim_fig} (right)
indicates that erroneous estimates almost never occur for $d_{\chi^2} \ga 15$. This behaviour is
generic and holds true for a wide range of model parameters (Fig. \ref{dchi2_vs_p1_fig}), suggesting that
the chi square distance is a useful parameter-free measure of distance between two Poisson intensities.
We thus select

\begin{equation}
d_{\chi^2}^{\rm max} = 50 \label{dmax} 
%d_{\chi^2} < d_{\chi^2}^{\rm max} = 50 \label{dmax} 
\end{equation}

as a safe cutoff for potentially successful candidate models. Candidate models which differ by more than $d_{\chi^2}^{\rm max}$ 
from the true model are not considered. The error introduced by this neglect is small; from the decay of the
histogram of best-fit distances $d_{\chi^2}$ we have estimated that less than $10^{-6}$ of all realizations are concerned.

\begin{figure}[h]
\centerline{\includegraphics[width=0.5\textwidth]{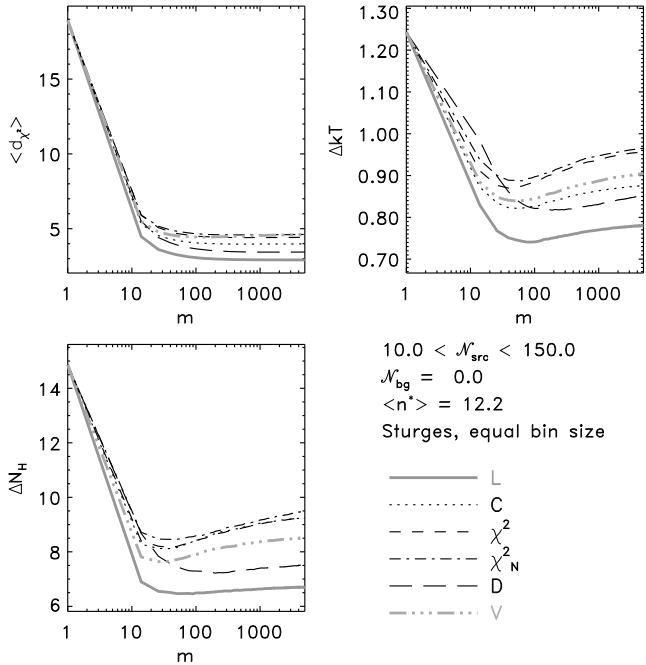}}
\caption{\label{explore_d_dens_fig}Similar to Fig. \ref{explore_d_diam_fig} but with $m$ proportional 
to the {\it density} of candidate models in parameter space.}
% created by ~unbinned/prog/revised/main2.f90 and result2.pro
\end{figure}

\subsection{Resolution}

In a second family of simulations, the restriction to the discrete models of Sect. \ref{source_sect} is relaxed
by interpolating the models in $(kT,N_H)$-space. Equation (\ref{dmax}) is then used to limit the search for best-fit
candidates. The number $m$ of candidate models thus delineates the {\it density} of models in parameter space,
and as $m$ is increased, the best-fit candidates converge to the best-fit solutions.
Figure \ref{explore_d_dens_fig} shows the result of this density exploration.
Note that the $\langle d_{\chi^2}\rangle$ curves converge at large $m$ 
to the same values as in Fig. \ref{explore_d_diam_fig}, confirming that $d_{\chi^2}^{\rm max}$ and 
the maximal $m$ are ample and do not effect the outcome of the simulation.

\begin{figure}[h]
\includegraphics[width=0.5\textwidth]{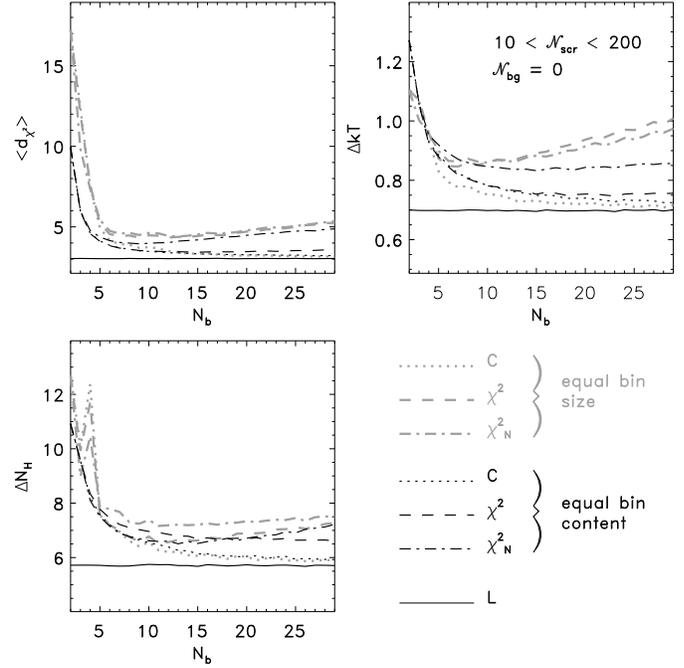}
\caption{\label{binning_fig}The effect of binning on the performances of ($C,\chi^2,\chi^2_{\rm N}$).
The corresponding unbinned $L$ values are shown for comparison.}
% created by ~unbinned/prog/unbinned3.f90 or unbinned3_bg.f90 and result3.pro
\end{figure}

\subsection{\label{binning_sect}Effect of binning, count rate, and background}

Up to here the binned methods invoked only Sturges' rule (Eq. \ref{sturges}) for better
comparability. In this Section, the effect of different binning methods and of the bin size is explored.
The case of a single bin must be excluded, since the predicted bin content 
is then insensitive to the parameters $kT$ and $N_H$.

In order to demonstrate the effect of the bin size we arbitrarily vary the number of bins $N_b$
(Fig. \ref{binning_fig}). Black curves refer to equal-content bins, and gray refer to 
equal-size bins. The curves represent averages over $(kT,N_H)$ and 10 $<$ $\N$ $<$ 200.
At small $N_b$ (many counts/bin), the ($C,\chi^2,\chi^2_{\rm N}$) curves collapse for both 
equal-content and equal-size cases. At large $N_b$ (few counts/bin), the performance of 
$C$ monotonically approaches the $L$ limit (black solid line). For large $N_b$, the $\chi^2$ 
and $\chi^2_{\rm N}$ statistics are not applicable due to low count rates. As a consequence,
their performance does not approach the $C$ and $L$ limit with increasing $N_b$.
This is especially pronounced for the equal-sized bins, which generally perform worse than equal-content bins.
From Fig. \ref{binning_fig} one may deduce the minimum number of counts 
per bin at which the chi square statistics apply: the increase of the ($\chi^2,\chi^2_{\rm N}$) curves at
$N_b \ga 10$ indicates that at least some 5-20 counts per bin should be present. By narrowing and shifting
the $\N$-cut one may explore the $\N$-dependence of the turnaround, and thereby find that 
$\chi^2$ and $\chi^2_{\rm N}$ require about 10 and 15 counts per bin, respectively. 

\begin{figure}[h]
\includegraphics[width=0.5\textwidth]{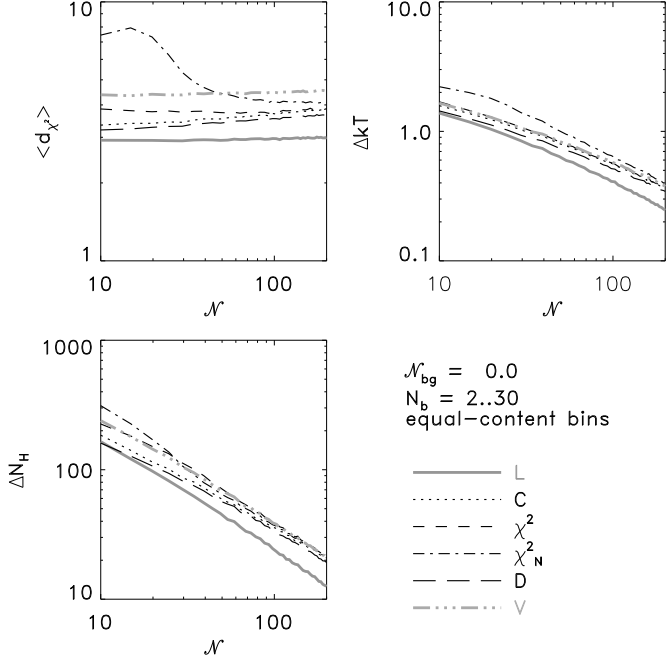}
\caption{\label{explore_N_fig}Dependence of the estimation errors on the true expected counts $\N$.}
% created by ~unbinned/prog/unbinned3.f90 and result3.pro
\end{figure}

Complementary to Fig. \ref{binning_fig}, Fig. \ref{explore_N_fig} shows the effect of $\N$, averaged
over $N_b$. Again, $N_b$ is independent of $\N$ for demonstration purposes. The bins have equal content. 
The $f$-space errors (top left) are found to be weakly dependent on $\N$ except for the $\chi^2_{\rm N}$ 
statistics, where the neglect of empty bins has a significant effect at low count rates ($\N \la 40$).
The errors $\Delta kT$ and $\Delta N_H$ scale like $\N^{-1/2}$, which is inherited from the
the normalization error $\Delta \N \sim \N^{1/2}$ (not shown).

We dismiss now the ad-hoc choice of $N_b$ used in Figs. \ref{binning_fig} and \ref{explore_N_fig}, and return
to the better adapted binning methods of Sect. \ref{bin_method_sect}. The number of counts per bin is 
thus determined from the Sturges' or AMISE rules, or fixed at $n^*$ = 8; the bins have either equal
size or equal (observed) content. The result of this simulation family is summarized in Table 
\ref{binning_tab1}, showing the performance ranking based on the mean chi square distance between 
true and estimated models. All simulations involve $10^5$ realizations, and the true 
source normalization is either $\N_{\rm src}$ = 20 or $\N_{\rm src}$ = 150, while the parameters
($kT,N_H$) are continuous (interpolated). A background of $\N_{\rm bg}$ = 50 
has been included for the sake of generality, so that both background-dominated and source-dominated
situations are addressed. From Table \ref{binning_tab1} it is seen that 
the exact likelihood ($L$) performs best and the Kolmogorov-Smirnov distances ($D$) performs worst. The 
relatively bad performance of $D$ compared to Sect. \ref{safe_discrim_sect} is found to be caused by the 
background. Pearson's $\chi^2$ generally performs
better than Neyman's $\chi^2_{\rm N}$; Kuiper's $V$ performs better than Kolmogorov's $D$.
A detailed comparison of pairs of simulations differing only by the count rate reveals
that the chi square distance between true and estimated models is always smaller for $\N$ = 150
than for $\N$ = 20, which is expected since more counts allow a sharper distinction of
the models. Equal-content bins generally perform better than equal-size bins. 

When the area under the operator-receiver curve (AUC; Sect. \ref{ROC_sect}) is used as an alternative performance measure 
instead of the average chi square distance between true and estimated models, we obtain the results of Tab. \ref{binning_tab2}.
This is similar to Tab. \ref{binning_tab1} but with $A \succ B$ indicating that $\langle {\rm AUC} \rangle_A < 0.99 \, 
\langle {\rm AUC} \rangle_B$ and $A \succsim B$ indicating
$0.99 \langle {\rm AUC} \rangle_B < \langle {\rm AUC} \rangle_A < \langle {\rm AUC} \rangle_B$
(like small $\langle d_{\chi^2} \rangle$, small $\langle {\rm AUC} \rangle$ indicates close agreement of
true and estimated models).
The average is over $10^4$ pairs of models covering the (continuous) parameter space ($kT,N_H$), with each
pair of models being probed by $2 \cdot 10^4$ realizations of Poisson data.
Although the AUC criterion is quite different in spirit from the chi square distance criterion
(binary classifier versus estimation problems), the performance orderings obtained by the two methods are 
surprisingly similar. In particular, the unbinned likelihood ($L$) performs always best, while the
internal performance ordering of ($C,\chi^2,\chi^2_{\rm N},D,V$) partially changes.
It was verified that $L$ performs best not only on the average, but also for all model pairs individually. 

\begin{table}[ht]
\begin{tabular}{c@{\hspace{2mm}}c@{\hspace{2mm}}r@{\hspace{2mm}}|c@{\hspace{2mm}}c@{\hspace{2mm}}c@{\hspace{2mm}}c@{\hspace{2mm}}c@{\hspace{2mm}}c@{\hspace{2mm}}c@{\hspace{2mm}}c@{\hspace{2mm}}c@{\hspace{2mm}}c@{\hspace{2mm}}c}
method & binning & $\N_{\rm src}$ & & & & & & \\\hline
Sturges & e.s. &  20.0  &  $L$ & $\succ$ & $\chi$ & $\succsim$ & $C$ & $\succsim$ & $\chi_{\rm N}$ & $\succsim$ & $V$ & $\succ$ & $D$ \\
Sturges & e.s. & 150.0  &  $L$ & $\succ$ & $V$ & $\succsim$ & $\chi$ & $\succsim$ & $C$ & $\succ$ & $\chi_{\rm N}$ & $\succsim$ & $D$ \\
Sturges & e.c. &  20.0  &  $L$ & $\succ$ & $C$ & $\succsim$ & $\chi$ & $\succsim$ & $\chi_{\rm N}$ & $\succ$ & $V$ & $\succ$ & $D$ \\
Sturges & e.c. & 150.0  &  $L$ & $\succ$ & $C$ & $\succsim$ & $\chi_{\rm N}$ & $\succsim$ & $\chi$ & $\succ$ & $V$ & $\succ$ & $D$ \\
AMISE & e.s. &  20.0  &  $L$ & $\succsim$ & $\chi$ & $\succsim$ & $C$ & $\succ$ & $V$ & $\succsim$ & $\chi_{\rm N}$ & $\succ$ & $D$ \\
AMISE & e.s. & 150.0  &  $L$ & $\succsim$ & $C$ & $\succ$ & $\chi$ & $\succ$ & $V$ & $\succ$ & $\chi_{\rm N}$ & $\succ$ & $D$ \\
AMISE & e.c. &  20.0  &  $L$ & $\succsim$ & $C$ & $\succ$ & $\chi$ & $\succsim$ & $\chi_{\rm N}$ & $\succ$ & $V$ & $\succ$ & $D$ \\
AMISE & e.c. & 150.0  &  $L$ & $\succsim$ & $C$ & $\succ$ & $\chi$ & $\succ$ & $\chi_{\rm N}$ & $\succ$ & $V$ & $\succ$ & $D$ \\
$n^*$=8.0 & e.s. &  20.0  &  $L$ & $\succ$ & $\chi$ & $\succsim$ & $C$ & $\succ$ & $\chi_{\rm N}$ & $\succ$ & $V$ & $\succ$ & $D$ \\
$n^*$=8.0 & e.s. & 150.0  &  $L$ & $\succ$ & $C$ & $\succsim$ & $\chi$ & $\succ$ & $V$ & $\succsim$ & $\chi_{\rm N}$ & $\succ$ & $D$ \\
$n^*$=8.0 & e.c. &  20.0  &  $L$ & $\succ$ & $C$ & $\succsim$ & $\chi$ & $\succsim$ & $\chi_{\rm N}$ & $\succ$ & $V$ & $\succ$ & $D$ \\
$n^*$=8.0 & e.c. & 150.0  &  $L$ & $\succsim$ & $C$ & $\succsim$ & $\chi$ & $\succsim$ & $\chi_{\rm N}$ & $\succ$ & $V$ & $\succ$ & $D$ \\
\end{tabular} 
\caption{\label{binning_tab1}Performance ordering of the statistics (Eqs. \protect\ref{L}-\protect\ref{V}), based on the average chi square 
distance between true and estimated models. The notation $A \succ B$ indicates here that 
$\langle d_{\chi^2} \rangle_A < 0.95 \langle d_{\chi^2} \rangle_B$, and $A \succsim B$ indicates that
$0.95 \langle d_{\chi^2} \rangle_B < \langle d_{\chi^2} \rangle_A < \langle d_{\chi^2} \rangle_B$.
The labels `e.s.' and `e.c.' refer to `equal size' and `equal count' bins, respectively. $\N_{\rm bg}$ = 50.}
% created by ~/manuel/unbinned/prog/revised/sim6.pro
\end{table}

\begin{table}[ht]
\begin{tabular}{c@{\hspace{2mm}}c@{\hspace{2mm}}r@{\hspace{2mm}}|c@{\hspace{2mm}}c@{\hspace{2mm}}c@{\hspace{2mm}}c@{\hspace{2mm}}c@{\hspace{2mm}}c@{\hspace{2mm}}c@{\hspace{2mm}}c@{\hspace{2mm}}c@{\hspace{2mm}}c@{\hspace{2mm}}c}
method & binning & $\N_{\rm src}$ & & & & & & \\\hline
Sturges & e.s. &  20.0  &  $L$ & $\succ$ & $V$ & $\succsim$ & $C$ & $\succsim$ & $\chi$ & $\succ$ & $\chi_{\rm N}$ & $\succ$ & $D$ \\
Sturges & e.s. & 150.0  &  $L$ & $\succ$ & $V$ & $\succsim$ & $C$ & $\succsim$ & $\chi$ & $\succ$ & $D$ & $\succsim$ & $\chi_{\rm N}$ \\
Sturges & e.c. &  20.0  &  $L$ & $\succ$ & $C$ & $\succsim$ & $\chi$ & $\succ$ & $\chi_{\rm N}$ & $\succsim$ & $V$ & $\succ$ & $D$ \\
Sturges & e.c. & 150.0  &  $L$ & $\succ$ & $C$ & $\succsim$ & $\chi$ & $\succsim$ & $V$ & $\succsim$ & $\chi_{\rm N}$ & $\succ$ & $D$ \\
AMISE & e.s. &  20.0  &  $L$ & $\succ$ & $C$ & $\succ$ & $\chi$ & $\succ$ & $V$ & $\succ$ & $\chi_{\rm N}$ & $\succ$ & $D$ \\
AMISE & e.s. & 150.0  &  $L$ & $\succsim$ & $C$ & $\succ$ & $\chi$ & $\succ$ & $V$ & $\succ$ & $D$ & $\succsim$ & $\chi_{\rm N}$ \\
AMISE & e.c. &  20.0  &  $L$ & $\succ$ & $C$ & $\succ$ & $\chi$ & $\succ$ & $V$ & $\succsim$ & $\chi_{\rm N}$ & $\succ$ & $D$ \\
AMISE & e.c. & 150.0  &  $L$ & $\succsim$ & $C$ & $\succ$ & $\chi$ & $\succ$ & $V$ & $\succsim$ & $\chi_{\rm N}$ & $\succ$ & $D$ \\
$n^*$=8.0 & e.s. &  20.0  &  $L$ & $\succ$ & $C$ & $\succ$ & $\chi$ & $\succ$ & $V$ & $\succ$ & $\chi_{\rm N}$ & $\succ$ & $D$ \\
$n^*$=8.0 & e.s. & 150.0  &  $L$ & $\succsim$ & $C$ & $\succ$ & $\chi$ & $\succ$ & $V$ & $\succ$ & $D$ & $\succsim$ & $\chi_{\rm N}$ \\
$n^*$=8.0 & e.c. &  20.0  &  $L$ & $\succ$ & $C$ & $\succ$ & $\chi$ & $\succ$ & $\chi_{\rm N}$ & $\succ$ & $V$ & $\succ$ & $D$ \\
$n^*$=8.0 & e.c. & 150.0  &  $L$ & $\succsim$ & $C$ & $\succ$ & $\chi$ & $\succ$ & $\chi_{\rm N}$ & $\succsim$ & $V$ & $\succ$ & $D$ \\
\end{tabular} 
\caption{\label{binning_tab2}Similar to Tab. \protect\ref{binning_tab1}, but using the area under 
the receiver-operator curve (Fig. \protect\ref{roc_fig}) as a measure of performance. $\N_{\rm bg}$ = 50.}
% created by ~/manuel/unbinned/prog/revised/sim7.pro
\end{table}

\begin{figure}[h]
\centerline{
\includegraphics[width=0.25\textwidth]{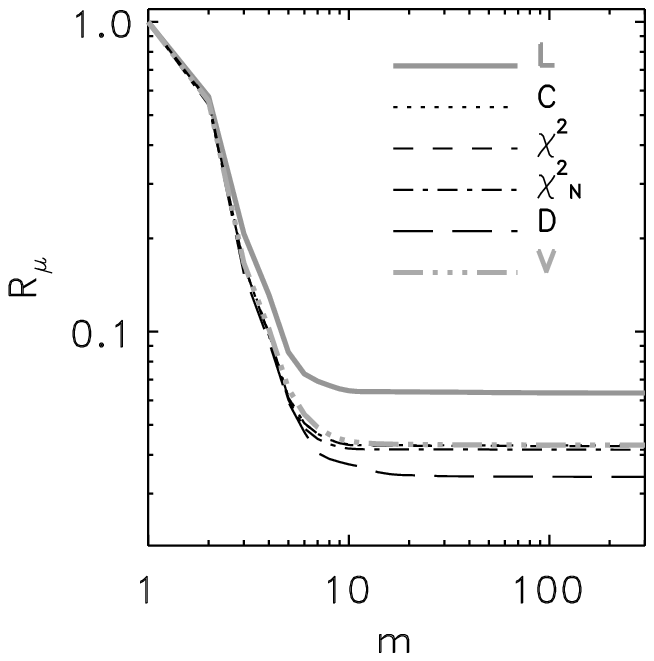}
\includegraphics[width=0.25\textwidth]{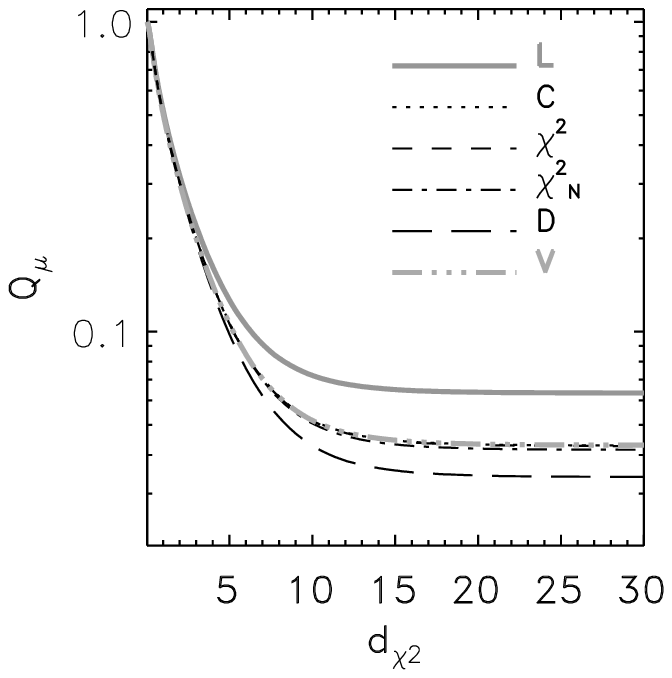}
}
\caption{\label{S_mu_sim_bg_fig}Similar to Fig. \protect\ref{S_mu_sim_fig}, but including background ($\N_{\rm bg}$ = 50).}
% created by ~unbinned/prog/unbinned1_bg.f90 and result1.pro
\end{figure}

Figures \ref{S_mu_sim_fig} - \ref{explore_N_fig} did not include background ($\N_{\rm bg} = 0$). 
In order to investigate the effect of the background we have repeated the simulations with variable 
background ratio $\N_{\rm bg}/\N$.
It turns out that the presence of a background does not affect superior performance of
the unbinned likelihood $L$. There are, however, differences in the performance ranking of the other statistics;
in particular, the unbinned $D$ and $V$ statistics are found to be degraded by the background.
We shall not show here the full diagnostic applied to the zero-background case, but restrict ourselves 
to an exemplary result.
Figure \ref{S_mu_sim_bg_fig} shows the number of successful identifications versus the
number of models at choice for $\N_{\rm bg}$ = 50, and is to be compared to Fig. 
\ref{S_mu_sim_fig}. The lead of the $L$ statistic is even more pronounced in the presence of background; 
the performance ordering derived from Fig. \ref{S_mu_sim_bg_fig} is $L \succ V \succsim C \succsim 
\chi^2 \succsim \chi^2_{\rm N} \succ D$.

\section{\label{realdata_sect}Real-data application}

\subsection{Observations and forward model}

After establishing the superior performance of the unbinned likelihood by Monte-Carlo simulations, we
apply it to actual \XMM~observations of the TMC. The data set (Tab. \ref{realdata_tab})
comprises 14 objects, and has been selected for low numbers of observed counts ($N_{\rm cnt}$) and simplicity of the
spectra so that parameterisation by ($kT,N_H,\N$) is adequate. Also, we have as far as possible avoided cases
where the background is not well known, and where the maximum-likelihood parameters lie on the boundary of 
the XSPEC parameter space, which would require a more careful analysis (Protassov et al. \cite{protassov02}).
`Bad' time intervals with increased background were omitted.

The source models are similar as in Fig. \ref{template_spectra_fig} (top) but adapted for the \XMM~instrumental 
response and background. The background spectrum is estimated according to the procedure of Sect. \ref{bkg_section}, setting
$\kappa$ = 0.05 to characterize the approximate \XMM/EPIC resolution. Only the PN detector of the \XMM/EPIC 
instrument is used. The abundances, relative to solar 
values (Anders \& Grevesse \cite{anders89}), are representative for highly active young stars
with inverse FIP effect (Telleschi et al. \cite{telleschi05}, Scelsi et al. \cite{scelsi05}, Garcia-Alvarez et al. \cite{garcia05}): He:1.0 -- C: 0.45 -- N: 0.788 
-- O: 0.426 -- Ne: 0.832 -- Mg: 0.263 -- Al: 0.50 -- Si: 0.309 -- S:  0.417--  Ar: 0.55 -- Ca: 0.195 
-- Fe: 0.195 -- Ni: 0.195. The total exposure time $T_{\rm exp}$ used for the analysis is in the order of a few 10 kiloseconds, 
and is given in Table \ref{realdata_tab}. The best-fit fluxes are converted into luminosities $L_X$ 
using XSPEC and assuming a distance of 140 pc to the TMC.

\begin{figure}[h]
%\centerline{\includegraphics[width=0.47\textwidth]{result_revised.eps}}
\centerline{\includegraphics[width=0.47\textwidth]{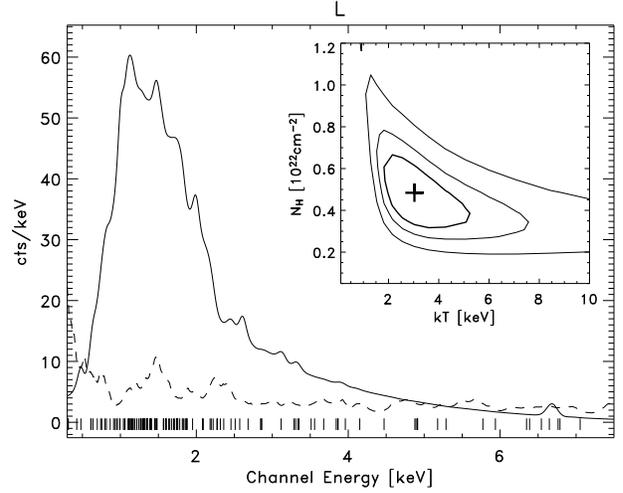}}
\caption{\label{Haro_6-13_fig}\XMM/EPIC observation of Haro 6-13, comprising 128 counts. Solid 
line: maximum-$L$ solution $f_{\rm src}(E)$; dashed: background spectrum $f_{\rm bkg}(E)$; ticks: 
observed counts. Inlets: best-fit parameters (crosses) and likelihood profiles (contours) at nominal confidence 
levels 68\% (boldface), 90\% and 99\%. See Sect. \protect\ref{confregion_sect}.}
% created by ~manuel/nbinned/prog/fit_xmm.pro
\end{figure}

\subsection{Best-fit models}

An example observation is shown in Fig. \ref{Haro_6-13_fig}, using data of Haro 6-13. 
The source extraction region contains 128 counts between 0.3 and 7.5 keV after elimination of 
bad time intervals, whereas the corresponding (scaled) background  contribution is $\N_{\rm bg}$ = 28.8 
counts. The observed counts are marked by ticks in Fig. \ref{Haro_6-13_fig}, while the background spectrum 
$f_{\rm bkg}(E)$ is shown by the dashed line. Energies below 0.3 keV are discarded. The maximum-likelihood
estimator for the source normalization is $\N_{\rm src}$ = 99.2 counts.
The unbinned log likelihood is computed inside the parameter cube
of Fig. \ref{Haro_6-13_fig} (inlet), and attains its maximum at $kT$ = 3.03
keV, $N_H$ = 0.48 $\times 10^{22}~{\rm cm}^{-2}$. These values are marked
by crosses in the inlets of Fig. \ref{Haro_6-13_fig}, and the corresponding best-fit model
$f_{\rm src}(E)$ is shown by solid line (main panel). For comparison, the
corresponding minimum-$\chi^2$ parameters, using 10 equal-content bins, 
are given by $kT$ = 2.55 keV, $N_H$ = $0.48 \times 10^{22}~{\rm cm}^{-2}$. The minimum 
$\chi^2$- and maximum $L$-estimates thus agree within errors (see below), but are not equal. The parameters 
$(kT,N_H)$ do not follow an equally simple trend, and are more sensitive to the statistics used. The other 
best-fit parameters listed in Table \ref{realdata_tab} have been obtained in a similar 
way as for Haro 6-13. 

\begin{table*}[ht]
\centerline{
\begin{tabular}{lccccccc}
%\input{table_revised.txt}
% copy-paste from "table_revised.txt" created by /afs/psi.ch/user/a/arzner/manuel/unbinned/prog/revised/collect_table.pro
Object	& $T_{\rm exp} [ks]$ & $N_{\rm cnt}$ & $kT$ [keV] & $N_H$ [$10^{22}{\rm cm}^{-3}$] & $\N_{\rm src}$ [counts] & $L_X$ [erg/s] \\\hline\\[-2mm]
             Haro 6-13 &  14.05 & 128 & $  3.03^{+  2.02\,(+  2.26)}_{ -0.88\,( -1.01)}$ & $  0.48^{+  0.12\,(+  0.13)}_{ -0.14\,( -0.15)}$ & $ 99.2^{+ 11.5\,(+ 11.4)}_{-10.7\,(-11.4)}$ & 1.5$\cdot 10^{29}$ \\[2mm]
            GN Tau ABC &  14.25 &  79 & $  4.57^{+  5.43\,(+  1.92)}_{ -2.48\,( -1.91)}$ & $  0.71^{+  0.71\,(+  0.35)}_{ -0.35\,( -0.24)}$ & $ 58.6^{+  9.1\,(+  8.0)}_{ -8.4\,( -9.6)}$ & 1.4$\cdot 10^{29}$ \\[2mm]
                TMC 1A &  14.25 &  39 & $  1.59^{+  5.82\,(+  4.87)}_{ -0.70\,( -1.49)}$ & $ 44.41^{+ 44.65\,(+ 15.44)}_{-30.94\,(-24.52)}$ & $ 24.0^{+  6.5\,(+  6.2)}_{ -5.9\,( -6.2)}$ & 4.7$\cdot 10^{30}$ \\[2mm]
       IRAS 04369+2539 &  14.25 & 148 & $  2.82^{+  4.30\,(+  3.84)}_{ -0.87\,( -1.54)}$ & $  6.51^{+  1.95\,(+  1.11)}_{ -2.40\,( -1.75)}$ & $120.4^{+ 12.0\,(+ 12.8)}_{-11.6\,(-11.3)}$ & 1.5$\cdot 10^{30}$ \\[2mm]
             HO Tau AB &  17.28 &  55 & $  0.14^{+  0.11\,(+  0.32)}_{ -0.09\,( -0.12)}$ & $  0.82^{+  0.36\,(+  0.20)}_{ -0.35\,( -0.63)}$ & $ 42.8^{+  7.7\,(+  7.7)}_{ -7.0\,( -7.3)}$ & 4.2$\cdot 10^{30}$ \\[2mm]
       IRAS 04325+2402 &  25.72 &  36 & $  2.70^{+  5.30\,(+  2.41)}_{ -1.70\,( -1.09)}$ & $  8.35^{+ 11.65\,(+  3.54)}_{ -5.27\,( -3.32)}$ & $ 21.6^{+  6.2\,(+  4.7)}_{ -5.6\,( -7.3)}$ & 2.1$\cdot 10^{29}$ \\[2mm]
      IRAS 04108+2803B &  26.76 & 467 & $  4.69^{+  1.41\,(+  2.01)}_{ -0.84\,( -1.10)}$ & $  7.73^{+  1.09\,(+  1.00)}_{ -0.95\,( -1.03)}$ & $380.8^{+ 21.3\,(+ 21.6)}_{-19.8\,(-22.4)}$ & 1.7$\cdot 10^{30}$ \\[2mm]
                CIDA 1 &  26.76 &  47 & $  0.78^{+  0.23\,(+  0.23)}_{ -0.57\,( -0.23)}$ & $  0.19^{+  0.68\,(+  0.18)}_{ -0.17\,( -0.27)}$ & $ 34.2^{+  7.1\,(+  5.9)}_{ -6.4\,( -7.9)}$ & 1.5$\cdot 10^{28}$ \\[2mm]
              V410 A13 &  37.27 &  26 & $  0.22^{+  0.56\,(+  0.44)}_{ -0.17\,( -0.14)}$ & $  1.14^{+  1.27\,(+  0.37)}_{ -0.89\,( -0.95)}$ & $ 12.7^{+  5.4\,(+  3.5)}_{ -4.7\,( -6.7)}$ & 5.1$\cdot 10^{29}$ \\[2mm]
      DD Tau AB$^{a)}$ &  37.26 & 395 & $  2.96^{+  0.92\,(+  1.10)}_{ -0.49\,( -0.56)}$ & $  0.40^{+  0.08\,(+  0.06)}_{ -0.06\,( -0.08)}$ & $325.8^{+ 19.6\,(+ 20.2)}_{-18.2\,(-19.5)}$ & 2.9$\cdot 10^{29}$ \\[2mm]
      DD Tau AB$^{b)}$ &  21.90 & 141 & $  2.40^{+  1.04\,(+  1.81)}_{ -0.51\,( -0.71)}$ & $  0.36^{+  0.16\,(+  0.09)}_{ -0.11\,( -0.14)}$ & $103.3^{+ 12.0\,(+ 11.9)}_{-11.2\,(-11.7)}$ & 1.4$\cdot 10^{29}$ \\[2mm]
V410 X6 &  37.23 & 158 & $  0.28^{+  0.41\,(+  0.18)}_{ -0.08\,( -0.09)}$ & $  0.74^{+  0.30\,(+  0.15)}_{ -0.50\,( -0.27)}$ & $122.2^{+ 12.7\,(+ 13.6)}_{-11.7\,(-11.6)}$ & 5.8$\cdot 10^{29}$ \\[2mm]
IRAS 04154+2823$^{a)}$ &  21.89 &  24 & $  2.66^{+  2.34\,(+  0.99)}_{ -1.66\,( -1.01)}$ & $  5.19^{+  2.81\,(+  1.00)}_{ -4.19\,( -3.19)}$ & $ 11.1^{+  5.2\,(+  4.5)}_{ -4.5\,( -5.8)}$ & 7.5$\cdot 10^{28}$ \\[2mm]
IRAS 04154+2823$^{b)}$ &  37.27 & 145 & $  5.09^{+  2.91\,(+  1.01)}_{ -2.62\,( -2.32)}$ & $  5.28^{+  2.98\,(+  2.05)}_{ -2.15\,( -1.06)}$ & $ 89.5^{+ 12.2\,(+ 10.6)}_{-11.5\,(-13.6)}$ & 2.3$\cdot 10^{29}$ \\[2mm]
\end{tabular}}
\caption{\label{realdata_tab}Maximum-$L$ parameters deduced from \XMM~observations of
the TMC. Superscripts a) and b) refer to different observations of the same object; $T_{\rm exp}$ is the exposure time, and $N_{\rm cnt}$ is the observed number of counts 
(corresponding to $\N_{\rm src} + \N_{\rm bg}$). Errors outside brackets represent projections of the 68\% likelihood
surfaces (Fig. \protect\ref{Haro_6-13_fig} inlets); errors in brackets are obtained from Monte-Carlo simulations
(Sect. \protect\ref{confregion_sect}). Luminosities $L_X$ refer to $N_H$ = 0 and the energy band from 0.3 to 10 keV.}
\end{table*}

\subsection{\label{confregion_sect}Confidence regions}

A rough estimate on the errors of the best-fit parameters can be obtained from the quantity 
$2 \, \Delta L \doteq 2(L_{\max}-L)$, which is asymptotically chi square distributed with the number of degrees of 
freedom equal to the number of model parameters. Here, $L_{\max}= \max_f L$ is the maximum achievable log
likelihood, which is associated with the full parameter space of Wilks theorem (Wilk, 1938).
By thresholding $2 \, \Delta L$ at the $\alpha$-quantiles of a chi square distribution with 2 degrees
of freedom, accounting for the parameters $(kT,N_H)$, and 1 degree of freedom, accounting for the parameter $\N_{\rm src}$
approximate confidence domains in parameter space are obtained. They are shown by solid contours in
Fig. \ref{Haro_6-13_fig}, representing cuts through the likelihood surfaces at confidence levels 68\% (boldface), 
90\%, and 99\%. The first errors given in Table \ref{realdata_tab} (outside brackets) represent projections 
of the 68\% confidence surfaces obtained in this way. Note that the absolute credibility of the best-fit solution 
(i.e. the value of $L_{\max}$) is not addressed by the statistic $2 \, \Delta L$, which
decouples the goodness-of-fit problem from the confidence domain problem.

\begin{figure}[ht]
%\centerline{\includegraphics[width=0.47\textwidth]{result_revised.eps}}
\centerline{\includegraphics[width=0.47\textwidth]{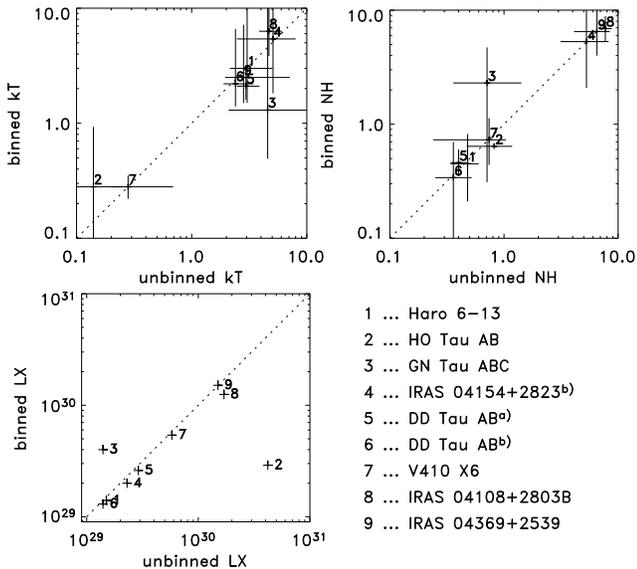}}
\caption{\label{comparison_fig}Comparison of the unbinned estimates with results from
binned one-temperature XSPEC fits. For $kT$ and $N_H$, the error bars refer to unbinned
68\% likelihood projections and XSPEC $\Delta \chi^2_{\rm N}$ = 1 levels (3 DOF).
For $L_X$, only the best-fit values are indicated.}
% this figure was created by /afs/psi.ch/user/a/arzner/manuel/unbinned/prog/show_comparison.pro
\end{figure}

\begin{figure*}[ht]
%\centerline{\includegraphics[width=0.4\textwidth,height=0.33\textwidth]{result_revised_2.0.eps}
%            \includegraphics[width=0.4\textwidth,height=0.33\textwidth]{result_revised_4.0.eps}}
\centerline{\includegraphics[width=0.4\textwidth,height=0.33\textwidth]{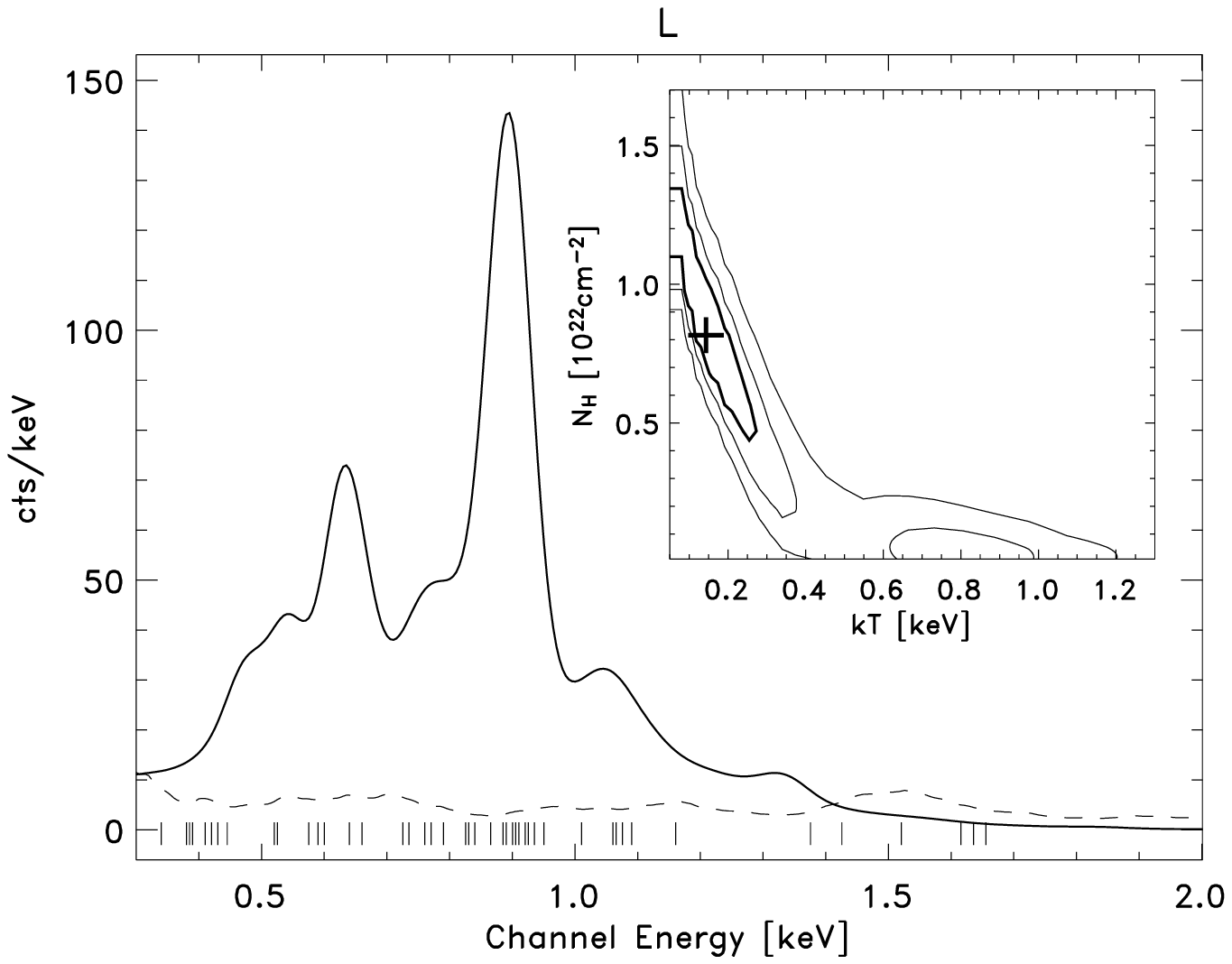}
            \includegraphics[width=0.4\textwidth,height=0.33\textwidth]{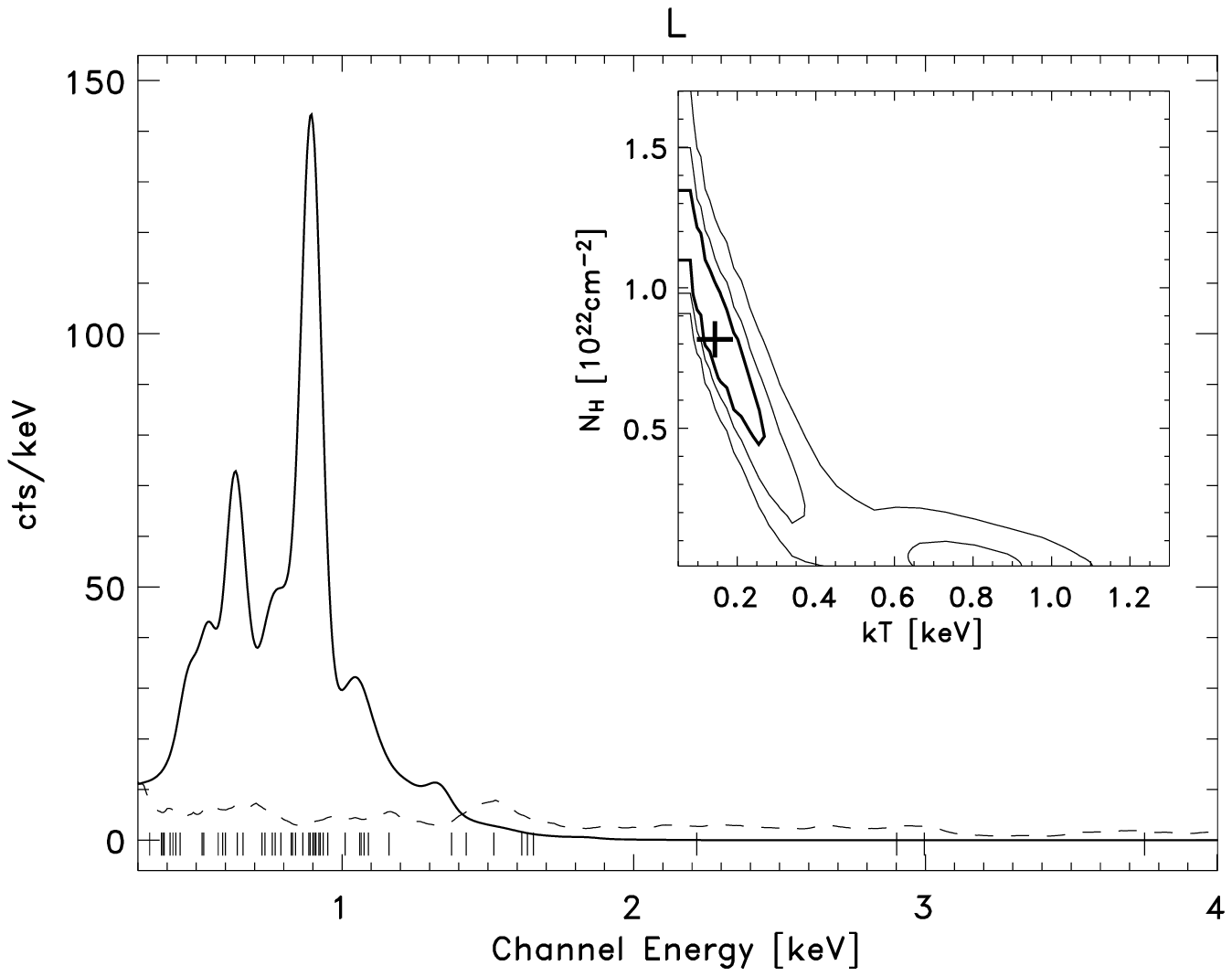}}
%\centerline{\includegraphics[width=0.4\textwidth,height=0.33\textwidth]{result_revised_chi2_2.0_5.eps}
%            \includegraphics[width=0.4\textwidth,height=0.33\textwidth]{result_revised_chi2_4.0_5.eps}}
\centerline{\includegraphics[width=0.4\textwidth,height=0.33\textwidth]{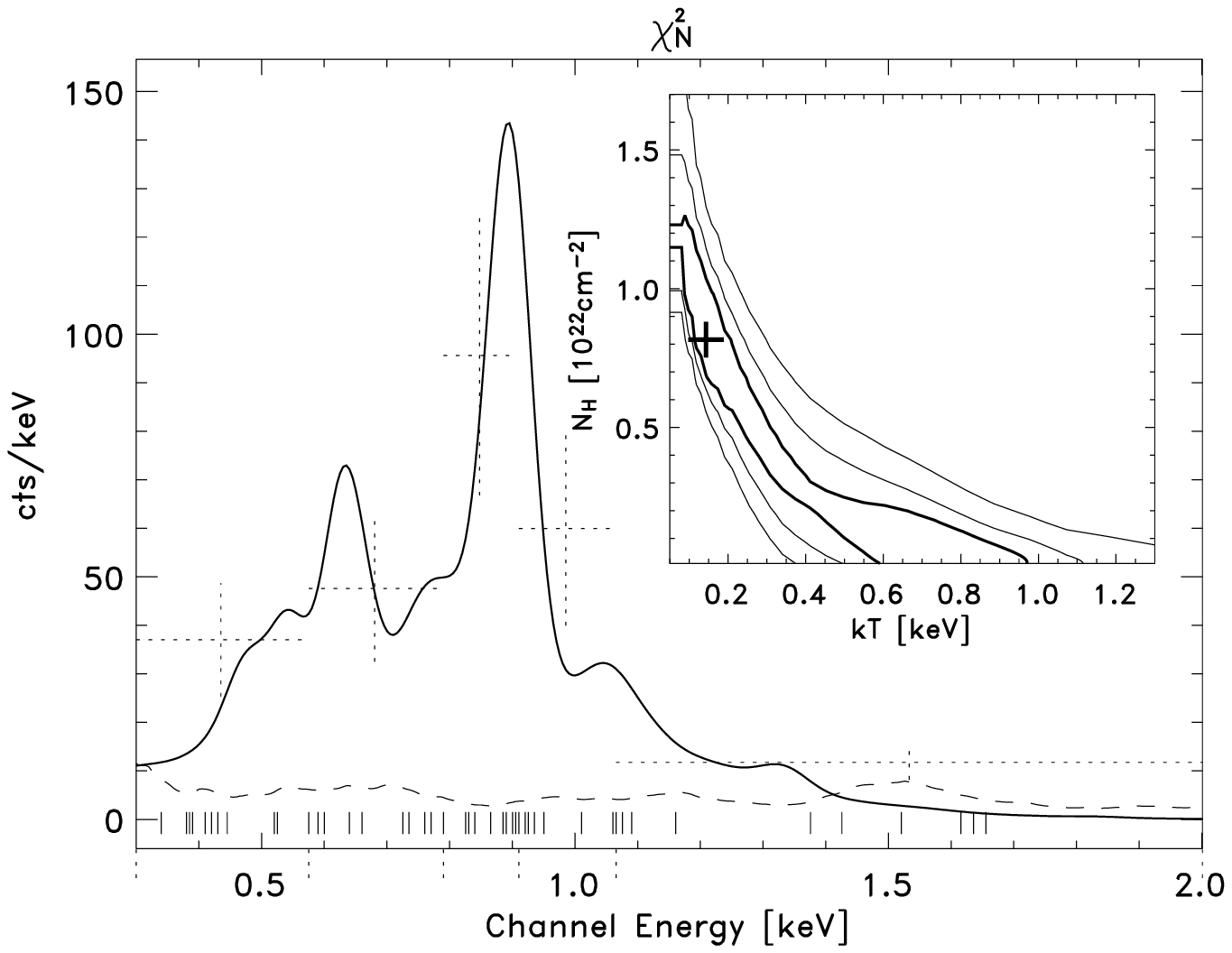}
            \includegraphics[width=0.4\textwidth,height=0.33\textwidth]{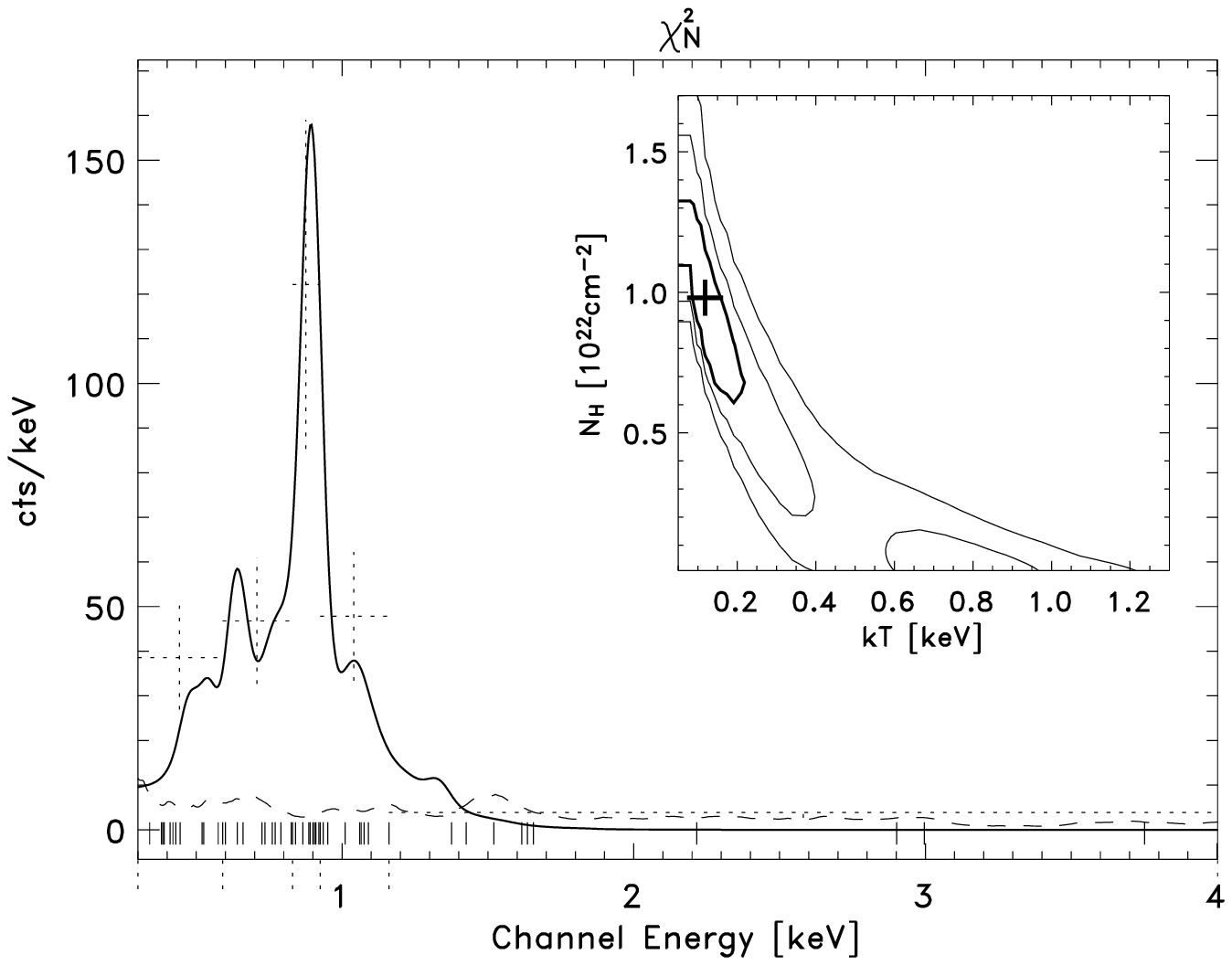}}
%\centerline{\includegraphics[width=0.4\textwidth,height=0.33\textwidth]{result_revised_chi2_2.0_4.eps}
%            \includegraphics[width=0.4\textwidth,height=0.33\textwidth]{result_revised_chi2_4.0_4.eps}}
\centerline{\includegraphics[width=0.4\textwidth,height=0.33\textwidth]{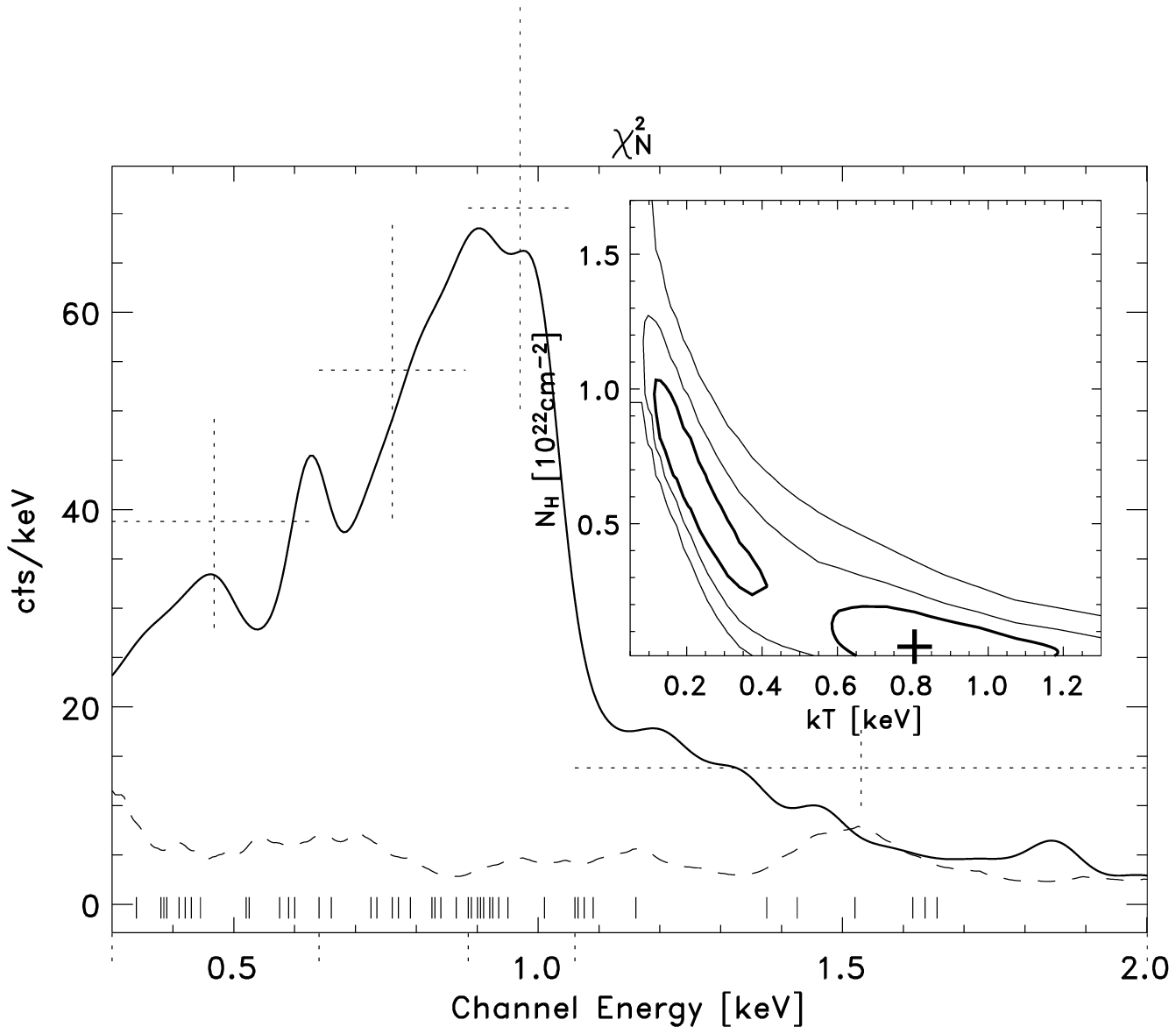}
            \includegraphics[width=0.4\textwidth,height=0.33\textwidth]{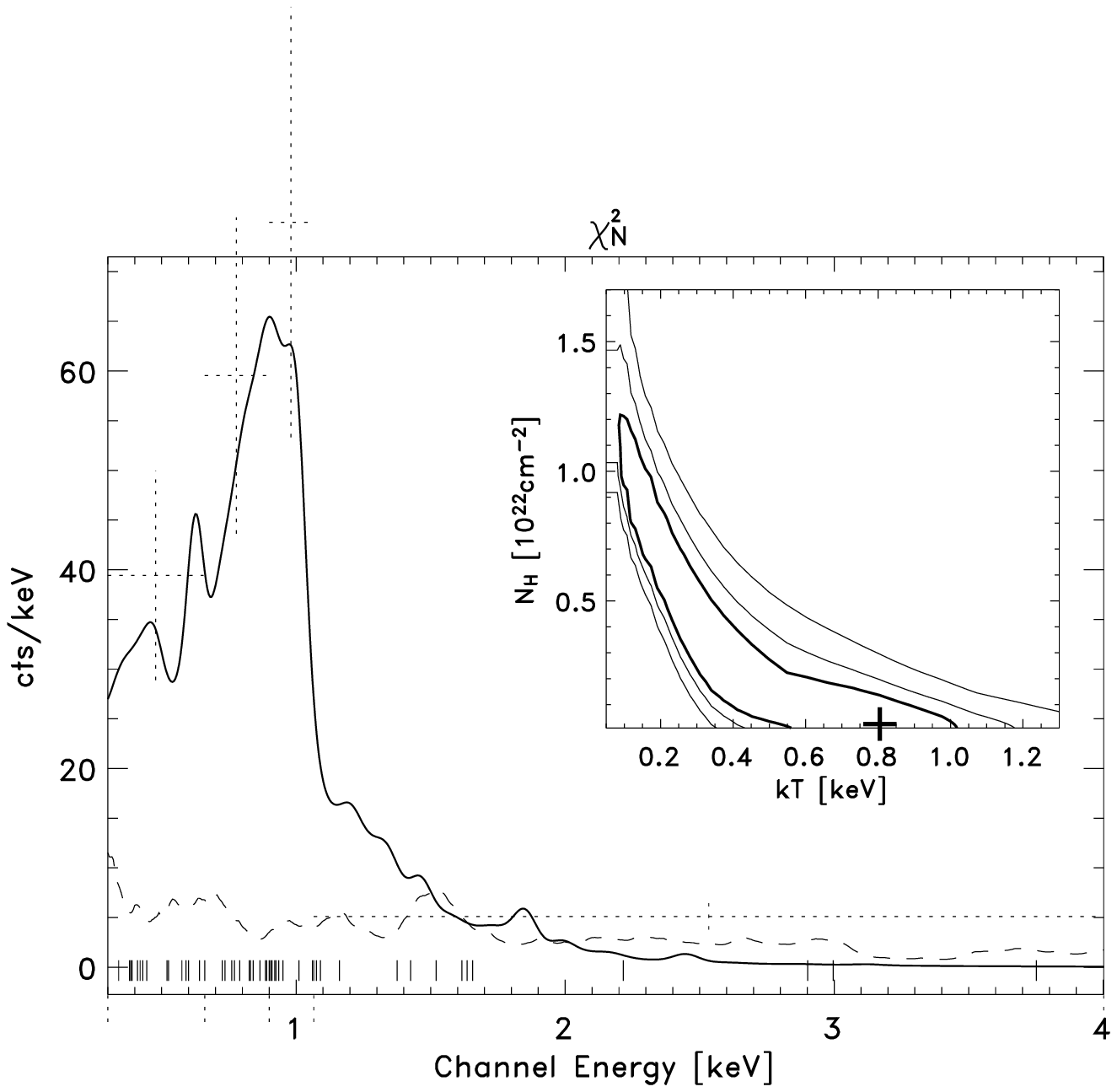}}    
\caption{\label{HOTau_result_fig}Exploration of the effect of energy band and binning on the HO Tau
parameter estimates. Left column: energies from 0.3 to 2 keV; right column: energies from 0.3 to 4 keV.
Top row: unbinned likelihood. Middle row: Neyman's chi square with 5 bins.
Bottom row: Neyman's chi square with 4 bins. The binned data are marked by dotted line.}
% eps files created by ~manuel/nbinned/prog/revised/fit_xmm.pro, and re-named according to Emax and Nbin
% data are from ../XMM/XEST-09-010/
\end{figure*}

The parameter errors can also be predicted from Monte-Carlo simulations alone, similar as in an instrumental design
study. To this end, the true and best-fit parameters (under $L$ statistic) of more than $10^6$ samples
are classified according to the best-fit parameters. For each class (containing about 100 samples),
the means ($b$) and standard deviations ($\sigma$) of the differences between true and best-fit parameters are
evaluated, and taken as a proxy for the biases and errors of the best-fit parameters. 
The simulation is repeated for each \XMM~observation, so that the correct background and instrumental response
are taken into account. The resulting error bounds $b \pm \sigma$ are given in parentheses 
in Table \ref{realdata_tab}, and are to be compared with the 68\% confidence limits from
the likelihood thresholding method. As can be seen, the Monte-Carlo errors are of the same order as
those obtained from the likelihood thresholding, and the biases are generally in consistent direction.

In summary, we have used here two complementary characterizations of parameter errors, based on
the log likelihood ratio $2 \, \Delta L$ and on Monte-Carlo simulations. The errors derived from the 
Monte-Carlo simulations do not invoke the asymptotic assumption underlying Wilks theorem, and are therefore 
better adapted to low count rates. But then, they do not involve the actual observation at all,
and rely entirely on the assumption that the true spectrum is correctly modeled by the template spectra
and the background model. Any systematic error contribution is thus neglected. Therefore, the errors from the 
Monte-Carlo simulation tend to under-estimate the true errors and should be considered as lower bounds. As
a further test, we have checked whether the minimum-$\chi^2$ estimates are within the errors of 
the maximum-$L$ solution, and found that this is so for all cases 
%except IRAS-04295+2251 
when $2 \, \Delta L$ 
is used, and holds true in half of the cases when the Monte-Carlo error bounds are invoked (both methods 
referring to 68\% confidence level).
% see /afs/psi.ch/user/a/arzner/manuel/unbinned/prog/fit_xmm.pro
We shall not pursue here a deeper discussion of confidence regions, but terminate with a hint to the literature
where more refined constructions may be found in Eadie et al. (\cite{eadie71}), Wachter et al. (\cite{wachter79}), 
Cousins (\cite{cousins95}), Porter (\cite{porter96}), Feldman \& Cousins (\cite{feldman98}), and Giunti (\cite{giunti99}). 
% ; a critical issue for classical Neyman confidence intervals of 
% binned Poison data is discussed in Arzner (Metrika, 2005)

\subsection{Comparison with binned estimates}

In order to compare the unbinned estimates with the standard XSPEC results, the XSPEC iterative
fitting package was applied to background-subtracted observations with more than 50 counts, using bins of (at least)
10 counts and default ($\chi^2_{\rm N}$) statistics. The spectral model is identical to the one used for unbinned estimation
(single temperature, $N_H$, abundances). The results are summarized in Figure \ref{comparison_fig},
displaying unbinned versus binned results. Numerical values for the binned results are tabulated G\"udel et al. (\cite{guedel06a}).
The crosses are centered at the best-fit solutions and
cover formal 1-$\sigma$ errors. For $L_X$, which is a derived quantity, only the best-fit parameters are indicated.
As can be seen, the agreement between binned and unbinned estimators is generally 
good, except for HO Tau and GN Tau.

In order to clarify the situation we have created plots of HO Tau similar to Fig. \ref{Haro_6-13_fig} for both 
the $L$ and $\chi^2_{\rm N}$ statistics, and have systematically varied the energy band and binning.
The binning procedure provides (approximately) equal number of observed counts per bin, running from low to high energies.
Some exemplary results are shown in Figure \ref{HOTau_result_fig}. The left column includes energies from 0.3 to 2 keV,
and the right column includes energies from 0.3 to 4 keV. The top row refers to the unbinned likelihood,
as quoted in Table \ref{realdata_tab} and Figure \ref{comparison_fig}.
The middle and bottom rows refer to Neyman's chi square statistic with 5 and 4 bins, respectively; containing 
(10,10,11,9,11), (11,11,11,11,11), (13,13,12,13), and (14,14,13,14) counts (fixed-$n^*$ method). The binned observed spectra are
indicated by the dotted crosses, with errors representing $\pm \sqrt{n}$ counts. As can be seen,
the unbinned likelihood has two shallow local maxima: a `cold' one at $kT$ $\la$ 0.2 keV and $N_H \sim 8 \times 10^{21}$ cm$^{-2}$, 
and a `warm' one at $kT \la 1$ keV and $N_H = 0$. While the (cold) maximum-$L$ solution and the $L$-profiles are stable under change  
of the energy range, the 68\% $\chi^2_{\rm N}$-profiles (middle) undergo a transition from a single
to two regions, and the minimum-$\chi^2_{\rm N}$ solution flips from cold to warm as the number of bins is
decreased (middle to bottom). This instability is mostly caused by the accumulation of counts around 0.8 keV, 
which either fall into a single (middle right) or two (bottom) bins. The 90\% $\chi^2_{\rm N}$ domain
is approximately stable under re-binning, and contains the XSPEC solution ($kT$ = 0.38 keV, $N_H = 6.4 \times 10^{21}\mbox{cm}^{22}$).
%(NB: the `warm' solution is also found by XSPEC if the `bad' time intervals are included.) 

Based on Figure \ref{HOTau_result_fig} and the instability of `warm' solutions under re-binning, one 
may conclude that the HO Tau data favour `cold' solutions, all the more so as the `warm' solutions lie 
on the border of the parameter space (Protassov et al. \cite{protassov02}). However, from an 
astrophysical point of view it is not obvious how such a cold and absorbed spectrum would
arise. A possible explanation is discussed in Section \ref{discussion_sect}. We shall therefore
argue that no definitive conclusion is possible for HO Tau yet, and that further observations are needed.

For the second discrepant case, GN Tau, the binned estimates yield smaller $kT$ and larger $N_H$ than
the unbinned ones (the binned estimates lie in the unbinned 90\% domain but outside the unbinned
68\% domain). This can be traced back to 12 observed counts below 1 keV; if these are excluded then the binned
and unbinned best-fit parameters agree within 20\%. We propose the following interpretation.
The presence of counts below 1 keV implies low absorption ($N_H$), provided that 
they cannot be explained by the background. The unbinned background model has, at $E$ $<$ 1 keV, fine-structures 
which do not coincide with the observed counts; hence the latter are attributed by the unbinned 
method to the source, entailing low $N_H$ and (Fig. \ref{template_spectra_fig} inlet) large $kT$. The binned 
method, in contrast, assigns more [namely, $\int_{0.3}^1 f_{\rm bg}(E) \, dE$] counts to the background; hence, $N_H$ is 
larger and $kT$ is smaller. The discrepancy of GN Tau stems thus from background fine structures at
energies below 1 keV.

\section{\label{discussion_sect}Summary and conclusions}

We have used Monte-Carlo simulations to assess the performance of the unbinned (exact) Poisson likelihood ($L$), 
binned Poisson likelihood ($C$), Pearson's $\chi^2$ and Neyman's $\chi^2_{\rm N}$, Kolmogorov-Smirnov 
($D$), and Kuiper's ($V$) statistics in the model classification and point estimation problems of low-count 
\XMM~and {\it Chandra}/ACIS spectra parameterized by temperature ($kT$), hydrogen density ($N_H$) and normalization
($\N$). By `unbinned' we mean that no grouping of instrumental energy channels is involved, so that the full readout 
resolution is used. The $L$ statistic equals the $C$ statistic on the finest possible (instrumental) binning.
In all cases, the source normalization $\N_{\rm src}$ was taken from its maximum-likelihood estimator, while the
shape parameters ($kT, N_H$) were taken such as to optimize the different statistics.

The outcome of the simulations can be summarized as follows:

\begin{itemize}

\item The unbinned Poisson likelihood performs best with regard to the following measures of performance:
(i) the probability of a successful identification in a search over discrete $d_{\chi^2}$-ordered candidate
models, (ii) the expected chi-square distance between continuously indexed true and estimated models, (iii)
the area under the operator-receiver curves of reduced binary (two-model) classifier problems, and (iv)
and generally also with respect to the mean square errors of individual parameter errors.

\item Under the $L$ statistic, two models can on average be distinguished if their chi square distance exceeds
$\sim$3. Similar statements hold for the other statistics ($C,\chi^2,\chi^2_{\rm N},D,V$) and correspondingly
larger distances (up to $\sim$5). The chi square distance is thus empirically found to be a useful parameter-free
measure of discrepancy between Poisson intensities. 

%\item The probability of accepting, under $L$ statistic, a model $f_2$ which is 
%$d_{\chi^2}(f_1,f_2)$ away from the true model $f_1$ decays exponentially with $d_{\rm \chi^2}(f_1,f_2)$.

\item The $\chi^2$ statistics should not be used unless more than 10 counts per bin are available, and $\chi^2_{\rm N}$ should not
be used for less than 15 counts.

\end{itemize}

We therefore argue that the unbinned Poisson likelihood $L$ is beneficial in cases with fewer
counts than instrumental readout channels. Using $L$, the narrow spectral lines and steep gradients 
are taken into account at their exact location, unaffected by the bin size.
Even more important, any instability of results under (arbitrary) choices of bin offsets and -sizes is avoided.
In our view, this is a major benefit since it eliminates user-dependent sources of discrepant results.

The $L$ statistics has been applied to XEST data, and demonstrated to operate
under real observational conditions. The major selection criteria
were spectral simplicity (so that the 3-parameter model applies), low count rate, and comparably 
well-known background. The parameter space was rigorously explored to avoid potential difficulties with
iterative optimization. The maximum-$L$ results usually agree with the minimum-$\chi^2$
and minimum-$\chi^2_{\rm N}$ values within 68\% confidence, but are not equal. There are, though, two cases (out of 14) 
where the unbinned and binned estimates disagree: HO Tau and GN Tau. 
Such disagreement, by itself, is a helpful indicator for potential problems in the forward modeling.

In particular, HO Tau is found by the $L$ method to be rather cool ($kT \sim$ 0.2 keV).
Such a cool X-ray-emitting plasma would be unusual for a T Tauri star, the
majority of which have average temperatures of $> 10$~MK. In the present survey
dominant plasma with similarly low temperature is only found in the low
temperature component of the spectrum of the jet-driving sources DG Tau A, GV Tau A, and DP Tau,
which may be due to shocks in the jet (G\"udel et al. \cite{guedel05}, G\"udel et al \cite{guedel06b}). However, DG Tau A also has
significant plasma at very high temperatures ($\sim 50$~MK) and HO Tau has
no known jet. A dominant low-temperature corona has been found on the
nearest classical T Tauri star, TW Hya, which is thus far unique in this
respect. Apparent high densities in this cool plasma suggest that
collisionally-ionized shocks in columns of material accreting onto the
star may generate the X-ray emission (Kastner et al. \cite{kastner02}; Stelzer and
Schmitt 2004). TW Hya is an unusually old classical T Tauri star,
approximately 10 Myr old. Using the Siess et al. (2000) stellar evolution
models, HO Tau appears also to be a relatively old classical T Tauri star,
at 8--9~Myr. Its mass accretion rate of $1.3 \times
10^{-9}$~M$_{\odot}$\,yr$^{-1}$ (White and Ghez 2001) is several times
higher than that of TW~Hya. Therefore it is plausible that HO~Tau may be
analogous to TW~Hya, in which case the bulk of its X-ray emission may be
generated by accretion shocks.

On the other hand, we can check whether the derived $N_H$ agrees with expectations
from measured optical and near-infrared extinctions $A_V$ and $A_J$. For general
interstellar matter, the conversions are $N_H \approx 2\times 10^{21}~A_V$~cm$^{-2}$
and $N_H \approx 5.6\times 10^{21}~A_J$~cm$^{-2}$ (see Vuong et al. 2002 and
references therein). The optical extinction for HO Tau is $A_V = 1.11-1.13$~mag
(Kenyon \& Hartmann \cite{kenyon95}, White \& Ghez \cite{white01}), and the near-IR
extinction is $A_J = 0.32-0.46$~mag (Kenyon \& Hartmann \cite{kenyon95}, Brice\~no et al.
\cite{briceno02}), thus implying $N_H = (1.8-2.6)\times 10^{21}$~cm$^{-2}$ under the assumption
of standard gas-to-dust ratios. These estimates are in better agreement with the
``hot'' solution (Fig. 15) although the error ranges are large. We also note
that significant deviations from the standard interstellar relation between extinction
and X-ray absorption are well known (e.g., in the jet-driving stars mentioned
above). Such deviations may point to an ``anomalous'' gas-to-dust ratio for example
as a consequence of dust evaporation.

We also note that both the binned and unbinned methods provide low $kT$ and high $N_H$.
The discrepancy is in $L_X$ (see Fig. \ref{comparison_fig}) which is a consequence of large uncertainties in
the spectral integration because most of the soft emission from the cool plasma is
strongly absorbed. Indeed, if the temperature estimate is changed from 0.14 keV (unbinned) to 0.28 keV
(binned) then $L_X$ drops by about a factor ten. The $L_X$ of the unbinned method is indeed rather high given
a stellar bolometric luminosity of this system of $L_* = 0.17~L_{\odot} = 6.5\times 10^{32}$~erg~s$^{-1}$. 
Adopting the usual X-ray saturation level of $L_X/L_* = 10^{-3}$ (e.g., Vilhu \& Rucinski \cite{vilhu83}),
we would expect a maximum $L_X = 6.5\times 10^{29}$~erg~s$^{-1}$, more in line
with the binned solution. 
The low count-rate of HO Tau unfortunately precludes a high-resolution
X-ray spectroscopic study to study density-sensitive triplet emission-lines 
from He-like ions of O, Ne and Mg, which suggest high densities in the TW Hya plasma.
As for GN Tau, $A_J$ = 1.17~mag (Luhman \cite{luhman04}), hence we expect $N_H = 6.6\times 10^{21}$~cm$^{-2}$,
very close to the result from the unbinned method.

Although the Monte Carlo simulations suggest that the maximum-$L$ estimates are closer to 
the true values when averaged over a (hypothetical) ensemble of similar observations, a
few words of caution are in order. The present simulations have assumed that 
the true spectrum is contained in the set of candidate spectra. If this is not the case, 
or if the Poisson process assumption is violated, then the use of the most binding $L$ statistics also
introduces the most severe misinterpretations. A major cause of flawed models stems from the background
estimation. We have assumed here that this estimation is perfect, and have not considered background
errors (e.g., Conrad et al. \cite{conrad03}), nor the more complicated problem of estimating the 
background spectrum jointly with the source spectrum. An imprecise background model may be
the cause for the discrepancy between binned and unbinned estimates found in GN Tau, where
the most relevant low-energy background (0.3 to 1 keV) contains 200 counts only, implying about 30\% statistical 
background uncertainty at resolution $\Delta E \sim$0.04 keV.
Furthermore, we have focused here on the choice of likelihood
functions for the point estimation and model classification problems. Accordingly, our
treatment of parameterization issues and confidence regions was rather crude. In particular,
we did not attempt to introduce any (Bayesian) a priori information other than implicit in
the choice of the forward models.

Finally, it should be pointed out that the unbinned approach is in principle not restricted to the one-dimensional 
spectrum problem considered in this article, and that the energy tags might be replaced by tuples of photon energy, arrival time, and 
position on the detector. However, the corresponding forward models would involve many more degrees of freedom, to the 
point where a simple maximum-likelihood principle might no longer be sufficient to uniquely determine the solution.
Additional information (such as a Bayesian prior) would then be required in order to regularize the problem.
Also, the more elaborated forward models would be more vulnerable to systematic errors (i.e., from CCD boundaries). 
In view of the (sparse) data we shall not pursue these issues.

\begin{appendix}

\section{Generation of non-homogeneous Poisson variates by the inversion method}

This Appendix describes the inversion method (Lewis \& Shedler \cite{lewis79}; 
Devroye \cite{devroye86}), one of the two methods used here to simulate event lists. The inversion method,
in the present implementation, relies on Theorem 1.4 (Chapter 6) of Devroye (\cite{devroye86}) stating 
that if $0< X_1 < X_2 < ...$ is a homogeneous Poisson process with unit rate function
and $\Lambda(x)$ is a non-decreasing function with $\Lambda(0)$ = 0 then
$0 < \Lambda^{-1}(X_1) < \Lambda^{-1}(X_2) < ...$ is a non-homogeneous Poisson process with
cumulative rate function $\Lambda(x)$. This follows from the fact that if $F$ is a continuous 
(cumulative) distribution with inverse $F^{-1}$ and $U$ is uniformly distributed in the unit interval [0,1],
then $F^{-1}(U)$ has (cumulative) distribution $F$. Our numerical
implementation makes use of the monotony of $\Lambda(x)$ and the fact that $X_{i+1}-X_i$ is
exponentially distributed, which allows a successive computation
of $\Lambda^{-1}(X_i)$. It proceeds as follows. Let a rate function $\lambda(x)$ be specified by a 
sufficiently resolved discrete version $\lambda_j$ with $j$ ranging from 0 to $N_c-1$, and 
define the discrete cumulative rate function by $\Lambda_0 = \tau \lambda_0$, $\Lambda_j = \Lambda_{j-1} + \tau \lambda_i$, 
with $\tau = \N^{-1} (N_c-1)$ and $\N = \sum_{j=0}^{N_c-1} \lambda_j$. 
Then, an non-decreasing bin number sequence $n_k \in [0,1, ..., N_c-1]$ of a non-homogeneous Poisson process 
with intensity $\lambda_j$ is obtained by the following pseudocode,

\begin{enumerate}
\leftskip0mm 
\item Initialisation:\\
\leftskip2mm $j = 0$\\ 
\leftskip2mm $k = 0$\\
\leftskip2mm $X = - \tau \ln U_0$

\item Iteration:\\
\leftskip2mm while $X < (N_c-1)$ do

\leftskip4mm while ($\Lambda_j < X$) and ($j < N_c$) do \\
\leftskip6mm $j = j + 1$\\
\leftskip4mm end do \\
\leftskip4mm $n_k = j$\\
\leftskip4mm $X  = X - \tau \ln U_k$\\
\leftskip4mm $\;\; k = k + 1$
 
\leftskip2mm end

\end{enumerate}

where $U_k$ are uniform random numbers in $[0,1]$, so that $-\tau \ln U_k$ is exponentially distributed. 
It was numerically verified that the $n_k$ obtained  in the above way are Poisson distributed with intensity $\lambda_{n_k}$.
% see ~/manuel/unbinned/prog/check_invmethod.pro

\begin{acknowledgements}
We would like to thank the International Space Science Institut (ISSI) in Bern, Switzerland, for logistic
and financial support during several workshops on the TMC campaign. This research is based on observations
obtained with \XMM, an ESA science mission with instruments and contributions directly funded
by ESA member states and the USA (NASA). A.T. and M.G. acknowledge support from the Swiss National Science
Foundation under grant Nr. 20-66875.
\end{acknowledgements}

\end{appendix}

\end{document}